\documentclass[fleqn, 10pt]{wlscirep}

\newcommand{\beginsupplement}{%
        \setcounter{table}{0}
        \renewcommand{\thetable}{S\arabic{table}}%
        \setcounter{figure}{0}
        \renewcommand{\thefigure}{S\arabic{figure}}%
     }

\title{Sequences of purchases in credit card data reveal life styles in urban populations}

\author[1, 2]{Riccardo Di Clemente}
\author[3]{Miguel Luengo-Oroz}
\author[4]{Matias Travizano}
\author[1]{Sharon Xu}
\author[5]{Bapu Vaitla}
\author[1, 6, 7*]{Marta C. Gonz\'alez}

\affil[1]{Massachusetts Institute of Technology,  Department of Civil and Environmental Engineering,  Cambridge,  02139,  MA,  USA}
\affil[2]{University College London,  The Bartlett Centre for Advanced Spatial Analysis,  London,  WC1E 6BT,  United Kingdom}
\affil[3]{United Nations Global Pulse,  46th St \& 1st Ave,  New York,  10017,  NY,  USA}
\affil[4]{GranData,  550 15th St. Suite 36C San Francisco,  94103,  CA,  USA}
\affil[5]{Harvard University,  Department of Environmental Health,   677 Huntington Avenue Boston,  02115,  MA,  USA}
\affil[6]{Department of City and Regional Planning,  Berkeley,  94720-1820,  CA,  USA}
\affil[7]{Lawrence Berkeley National Laboratory,  1 Cyclotron Road,  Berkeley,  94720-1820,  CA,  USA}

\affil[*]{martag@mit.edu}

\begin{abstract}
Zipf-like distributions characterize a wide set of phenomena in physics,  biology,  economics and social sciences. In human activities,  Zipf-laws describe for example the frequency of words appearance in a text or the purchases types in shopping patterns. In the latter,  the uneven distribution of transaction types is bound with the temporal sequences of purchases of individual choices. In this work,  we define a  framework using a text compression technique on the sequences of credit card purchases to detect ubiquitous patterns of collective behavior. Clustering the consumers by their similarity in purchases sequences,  we detect five consumer groups. Remarkably,  post checking,  individuals in each group are also similar in their age,  total expenditure,  gender,  and the diversity of their social and mobility networks extracted by their mobile phone records. By properly deconstructing transaction data with Zipf-like distributions,  this method uncovers sets of significant sequences that reveal insights on collective human behavior. 
\end{abstract}
\begin{document}

\flushbottom
\maketitle
%
%
\thispagestyle{empty}

\section*{Introduction}

In the age of information,  we leave digital traces of our everyday activities: the people we call,  the places we visit,  the things we eat and the products we buy. Each of these human activities generates data that when analyzed over long periods and yields a comprehensive portrait of human behavior \cite{eagle2010network, giles2012making, lazer2009life, mervis2012agencies, pentland2013data, vespignani2012modelling}.

In the last decade,  Call Detailed Records (CDRs) have been of paramount importance to understand the daily rhythms of human mobility \cite{blondel2015survey, gonzalez2008understanding, jiang2016timegeo, song2010limits, toole2015coupling}. By properly analyzing billions of digital traces,  our modern society has a whole framework to analyze wealth \cite{blumenstock2015predicting},  socio-demographic characteristics \cite{lenormand2015influence},  and to better tackle the origins of urban traffic \cite{ccolak2016understanding, louail2014mobile}. 
By contrast,  we still need to better exploit Credit Card Records (CCRs) to uncover the behavioral information they may hide.  Main uses of CCRs have been to measure similarity in purchases via affinity algorithms \cite{pennacchioli2014retail, solomon2014consumer}.
Recent research has also shown that credit card data can be used analogously to mobile phone data to detect human mobility. Namely,  the CCRs inform us about the preferred transitions between business categories,  identifying the unevenness of the spatial distributions of people's most preferred shopping activities \cite{yoshimura2016urban},  and to enrich urban activity models. Consumers' habits are shown to be highly predictable \cite{krumme2013predictability},  and groups that share work places have similar purchase behavior \cite{dong2018social}. These results let possible to define spatial-temporal features to improve the estimates of the individual's financial well-being \cite{singh2015money}.

It has been measured by individual surveys and confirmed by credit card and cash data that the vast majority of daily purchases is dominated by food,  then followed by mobility and communication-social activities \cite{lenormand2015influence, matheny2016state}. Their frequency seems to follow a Zipf distribution,  meaning that the most frequent category of purchases will occur approximately twice as often as the second most frequent category,  three times as often as the third etc. Grouping the consumers depending on socio-demographic attributes preserves the zipf-like behavior and the dominant purchase (food). For each group,  there is a peculiar order in abundance of less frequent category. As pointed out by \cite{lenormand2015influence, sobolevsky2016cities} this depends on socio-demographic features such as income,  gender and age.

 Hence,  the challenge at hand is to obtain meaningful information within these highly uneven spending frequencies to capture a comprehensive picture of their shopping styles related with city socio-economic dynamics within the city. 
  
\vskip5pt
A similar challenge appears in the sequence of diseases in medical records \cite{roque2011using} or phenotype associations with diseases \cite{hidalgo2009dynamic}. Existing approaches cluster patients based on their historical medical records described by the International Classification of Diseases (ICD). In this case,  the frequency-inverse document frequency (TF-IDF) ranking is used to eliminate redundant information. 

In the matter of the uneven word frequency in the text corpora \cite{piantadosi2014zipf}. Bayesian inference methods have been used to detect the hidden semantic structure. In particular,  the Latent Dirichlet Allocation (LDA) \cite{blei2003latent} is a widely used method for the detection of topics (ensemble of words) from a collection of documents (corpus) that best represent the information in datasets.

However,  both of the above-mentioned approaches do not take into account the temporal order in the occurrence of the elements. Our goal is to eliminate redundancy while detecting habits and keeping the temporal information of the elements,  which in the case of purchases are an important signature of an individual's routine and connect them to their mobility needs.  In this work,  we identify significant ordered sequences of transactions and group the users based on their similarity. This allows to offer deeper description of consumers behavior unraveling their routines.

In this work,  we are interested in uncovering diverse patterns of collective behavior extracted from this data.  Specifically,  how the digital footprint of CCRs can be used to detect spending habits,  reflecting interpretable lifestyles in the population at large. By integrating credit card data,  with demographic information and mobile phone records,  we have a unique opportunity to tackle this question. 

 The presented method is able to deconstruct Zipf-like distribution into its constituents' distributions,  separating behavioral groups. Paralleling motifs in network science \cite{milo2002network},  which represent significant subnetworks,  the uncovered sets of significant sequences are extracted from labeled data with Zipf-type distribution. Applied to CCRs,  this framework captures the semantic of spending activities to unravel types of consumers. The resulting groups are further interpreted by coupling together their mobile phone data and their demographic information.  Consistently,  individuals within the five detected groups are also similar in age,  gender,  expenditure and in their mobility and social network diversity.  We show that the selection of significant sequences is a critical step in the process,  it improves on TF-IDF method that are not able to discern the spending habits within the data. Remarkably,  our results are comparable with the ones obtained by LDA,  with the added advantage that it takes into account the temporal sequence in the activities.

\section*{Results}
\subsection*{Data analysis}

\begin{figure}[!htb]
\centering
\includegraphics[width=.87\linewidth]{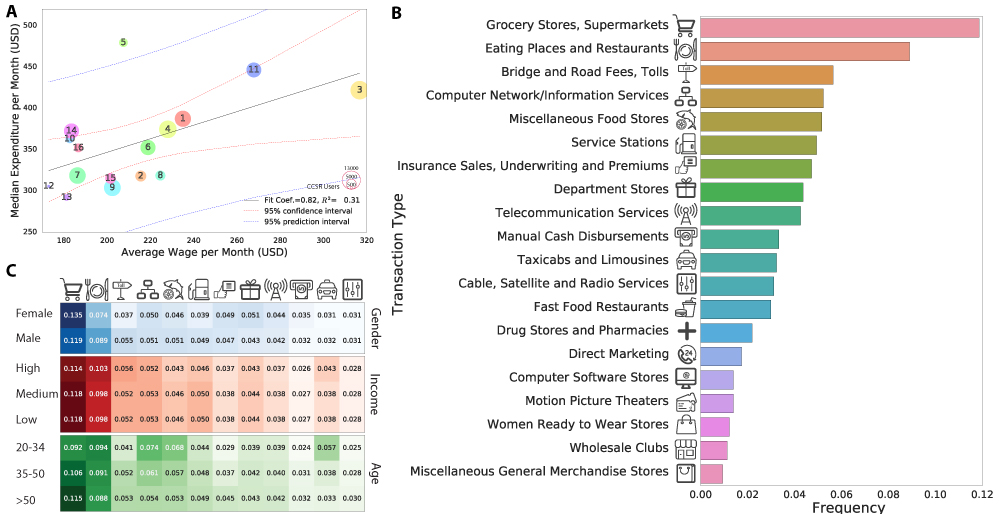}
\caption{\textbf{Transaction frequency by type and their demographics. (A)} User median expenditure per month in CCRs transactions vs. the average monthly wage in their district of residence. The color and the numbers represent different districts of Mexico City (see Supplementary Figure 1) and the size of the circles is proportional to the number of users in the district. \textbf{(B)} Transactions by type as define by MCC (17). \textbf{(C)} Comparison of frequencies by transaction types (same as in B) separating users in groups according to their gender,  income and age. The share of transaction frequency is distributed similarly among different groups. The icons used in this figure are work of Azaze11o/Shutterstock.com.}
\label{fig:fig1}
\end{figure}

We analyze individual CCRs transactions over 10 weeks of 150,000 users who live in one of the most populated cities in Latin America (Mexico City,  MEX). The dataset contains age,  gender,  and residential zipcode of the users (Supplementary Figure 1A-C).  For each user,  we analyze the chronological sequence of their transactions and the associated expenditure labeled with transaction type via a Merchant Category Code (MCC) \cite{solutions2004merchant}. The purchase entries are aggregated by user and are temporal ordered respect each day. For one tenth of the analyzed users we also have their CDR data over a period of 6 months (overlapping the CCRs time period),  including time,  duration,  location of the calls and id of the receiver. While the payment with cards and electronic payment terminals are being promoted in the region to improve financial inclusion,  credit card adoption rates remain relatively low at 18\% of the population \cite{pymnts2017}. First,  we check how representative the CCRs users are within the city. We observe the correlation between the median CCRs expenditure in the dataset at district level,  and the average monthly wage in the same district according to the census (Fig. \ref{fig:fig1}A) (Source: INEGI, National Survey of Occupation and Employment (ENOE). Population aged 15 years and older.). The monthly expenditure of the card users is high in relation to the monthly wages,  indicating that the adoption of credit cards predominantly occurs among users with higher wages in each district. However,  our users' sample spans over all the city districts with different income levels. We observe that the wider adoptions of credit card is across male and young adults (age 35-50) in each district (Supplementary Figure 1B-D-E-F).  The spending patterns in the CCRs reveal that the frequency of the purchase types follows a Zipf's law (Supplementary Figure 2A). The majority of shoppers use more frequently the top-twenty transactions codes presented in Fig. \ref{fig:fig1}B,  among hundreds of possible MCCs. Moreover,  slight variations emerge in this trend when dividing the population by wealth,  age and gender (Fig. \ref{fig:fig1}C). In general,  transaction codes related to food,  mobility and communication,  in that order,  dominate the number of top transactions in all groups,  and the number of transactions per day,  for each user,  is not affected by any socio-demographic category (Supplementary Figure 2B-C).

\subsection*{Credit card transactions codes as sequence of words}
Our main goal is to amplify the signal in the data to identify individuals' expenditure habits hidden in the non-uniform distribution of transaction types present in the Zipf's type of distribution. The first step in this direction is to transform the chronological sequence of user MCC codes into a sequence of symbols given by the transaction codes (Fig. \ref{fig:fig2}A). We apply the Sequitur algorithm \cite{nevill1997identifying} to infer a grammatical rule that generate words,  defined as MCC symbols that repeat in sequence. The result of this process applied recursively,  is a compression of the original sequence with new symbols called words,  which offer insights into the repeated sequences of transactions. We take each word as a routine in shopping as they are a chronological sequence of two or more MCCs that appear frequently. We detect more than ten thousand different words also following a Zipf type distribution as presented in Figure \ref{fig:fig3}. We notice that the inter-time transactions between in word purchases are smaller respect two random consecutive transactions. Moreover,  the time to perform a $n$-transactions word,  defined as the time between the first and the last purchase of the word,  it is smaller than the time of two consecutive transactions picked randomly (Fig. \ref{fig:fig3}C). The set of words $\{w_i\}$ for user i are significant only if their occurrence differs from the outcome of a random process with the same number of transactions per type. To detect the words that are significant,  we generate 1000 randomized code sequences for each user. For each realization,  we apply the Sequitur algorithm to define the words in the randomized sequences and evaluate the significance level of the user's words by computing the z-score of the occurrence of the real words with respect to the randomized ones. Z-score test needs to be performed on a Gaussian distribution of the word occurrence. The words occurrence distribution of simulated samples has in general a normal shape. But in several cases the frequency of the generated words has a small number of occurrence,  in Supplementary Figure 3-4 we show the robustness of a z-score benchmark to assess the word significance either for non-Gaussian distributions. We extract for each user the set of significant words with z-score greater than 2,  defined as $\{W_i\}$. The selected words represent the shopping routines that indicate informative choices in the user's spending behavior (see Supplementary Figure 5),  given their occurrence vary from the mean by two standards deviations. In the Supplementary Figure 5D-E we analyze the number of validate users' with at least a significant word depending on the z-score threshold.

\subsection*{The life styles}
\begin{figure}[!htb]
\centering
\includegraphics[width=.87\linewidth]{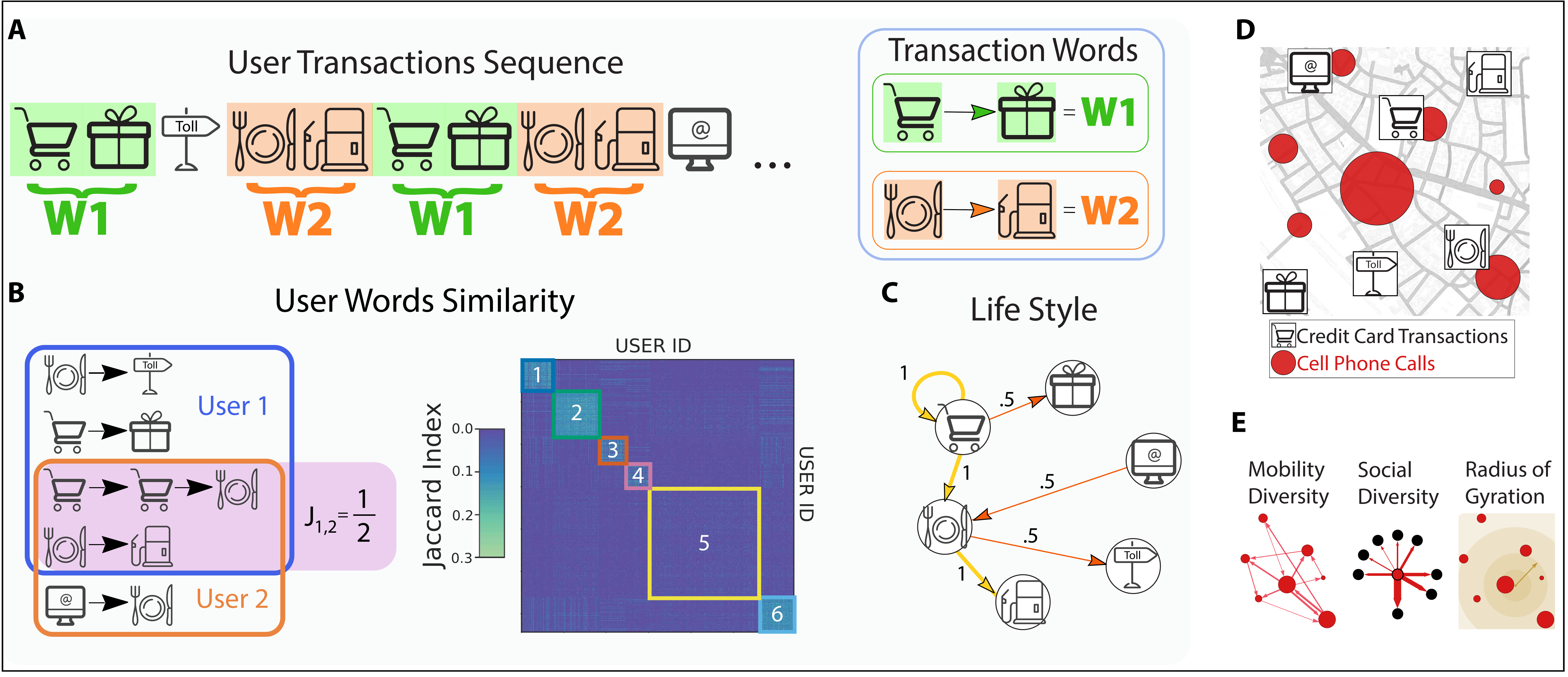}
\caption{\textbf{Methods and metrics (A)} Schematic representation of the Sequitur's algorithm applied to a sequence of transactions of one user to detect words and identify the significant transaction sequences in the dataset.  \textbf{(B)} Calculation of the similarity between two users (left) based on the Jaccard index of their significant sequences to define the matrix of users' similarity (right). Group of users are detected based on similar sequences of transactions.  \textbf{(C)} Life style representation based on sample users 1 and 2 of Fig. 2B. \textbf{(D)} Example of traces of CDR and CCRs data for the user. \textbf{(E)} Metrics adopted for the analysis of CDR data. The icons used in this figure are work of Azaze11o/Shutterstock.com.}
\label{fig:fig2}
\end{figure}

\begin{figure}[!ht]
\centering
\includegraphics[width=\linewidth]{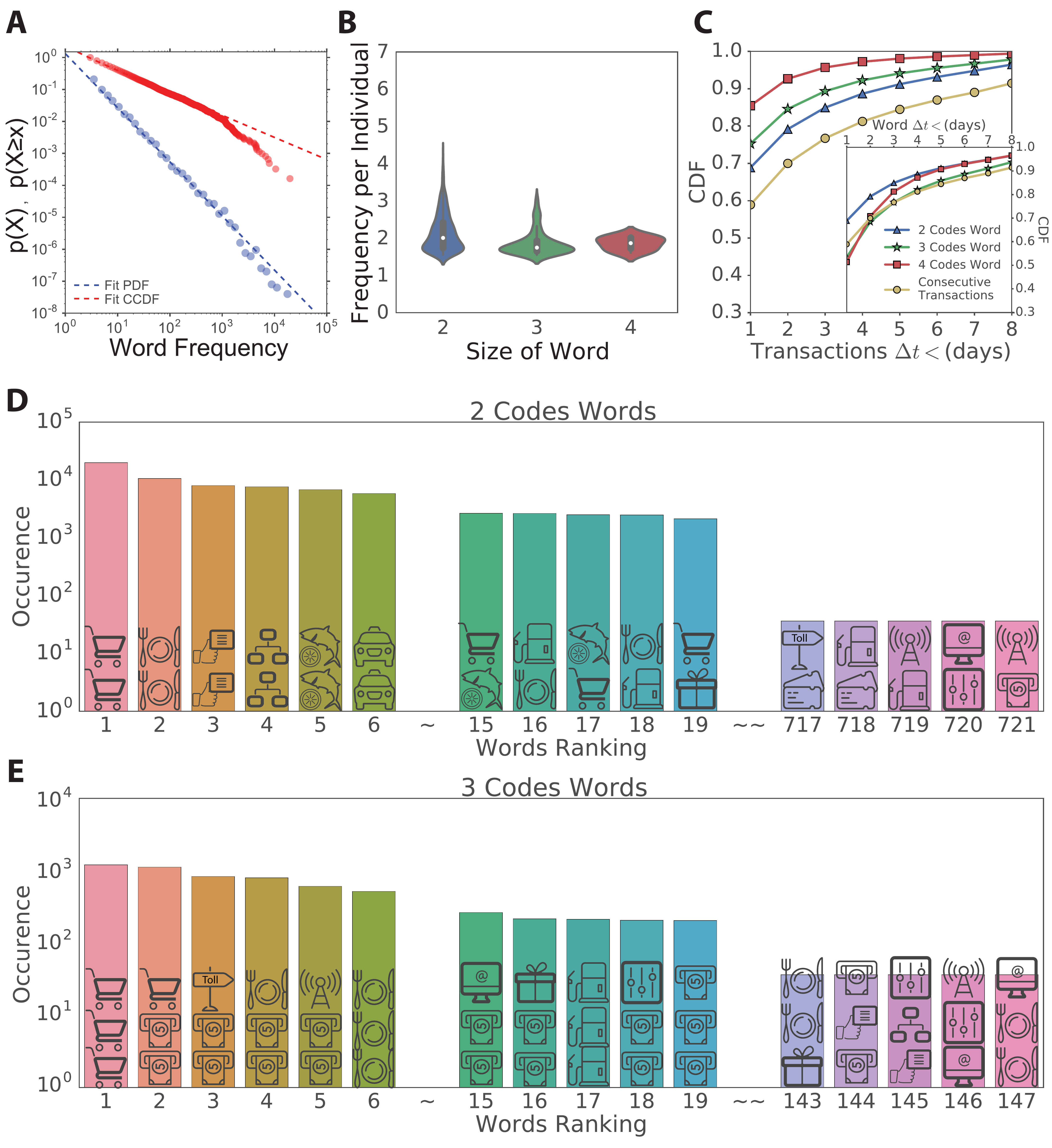}
\caption{\textbf{Semantic analysis of transaction sequences (A)} Probability Density Function plot of the occurrence of the words $\{w_i\}$ and its complementary cumulative distribution; the probability distribution words manifest a power law behavior $p(w_i)\propto x_i^{(-1.70)}$; with $x_i$frequency of the $\{w_i\}$ and Kolmogorov Smirnov distance $D_n=0.014$. \textbf{(B)} Distribution of the occurrence of words in transaction sequences by the word length. \textbf{(C)} Inter-time transactions between purchases. The purchases within each word are more likely to occur within a day respect two random consecutive transactions. \textbf{(C inside)} Moreover,  the purchase-time to accomplish a word completely is less respect two random consecutive transactions. \textbf{(D-E)} Examples of words composed by two and three codes respectively ordered by number of occurrences. The icons used in this figure are work of Azaze11o/Shutterstock.com.}
\label{fig:fig3}
\end{figure}

With this meaningful samples,  we can now measure the similarity between the shopping behavior among users. To that end,  we decompose each significant word as directed links between its transaction codes. Each user is represented by a directed network,  in the space of MCC,  that collects all the links present in the user's words. We then calculate the Jaccard similarity coefficient between all the users to compare the set of links in their networks (see the illustration of the method in Fig. \ref{fig:fig2}B). Since the users' networks have a low degree,  our similarity measure is not sensitive to the sets' size (Supplementary Figure 6C). Moreover,  our results are in agreement to the ones that use the turnover component of Jaccard dissimilarity index \cite{baselga2012relationship},  that is less susceptible to the sets' size (see Supplementary Figure 6). Thanks to the Jaccard Index we obtain the matrix M,  of users' similarity in shopping sequences.

Finally,  we identify groups in this matrix by applying a parallel Louvain algorithm to fast unfolding the communities in $M$ \cite{blondel2008fast, staudt2016engineering}. The same clusters appear with the Leading Eigenvector \cite{newman2006finding} and Walking Trap \cite{pons2005computing} (Supplementary Figure 7, 8).We detect six clusters or groups of users who share similarities in their spending habits,  one of the six encloses unlabeled users that are close to the average behavior,  while the other five present interesting behavioral preferences as confirmed later by their demographics and their mobile phone records.

Figure \ref{fig:fig2}C shows the group's shopping habits. The weight of the arrows between two codes represents the fraction of users of a given cluster that have the given transaction sequence. This schematic representation of the groups routines is possible because our method firstly,  detects the most significant sequences of transactions and secondly preserves the temporal information embedded in the word as the ordered sequence of transaction. 

\subsection*{Coupling Credit Card data with the Mobile Phone data}
In order to gather a more comprehensive portrait of the users' behavior,  we couple the information of the CCRs users with their CDR data (Fig. \ref{fig:fig2}D-E). From the mobile phone data,  we analyze basic characteristics of an individual's social contacts and her mobility network with well-established metrics. Namely,  the social diversity,  the homophily,  the mobility diversity,  the radius of gyration \cite{pappalardo2015using, gonzalez2008understanding},  the tower residual activity \cite{toole2012inferring} and the mobility behavioral pattern. The social network diversity is the entropy associated with the number of individual i's communication events with their reciprocal contacts divided by the number of contacts \cite{eagle2010network}. The homophily,  in the call graph from the mobile phone data,  is a metric that investigate whether or not two users in the same cluster have a higher probability of contacting each other. The mobility diversity is measured via the entropy in the number of trips between locations normalized by the number of visited locations \cite{pappalardo2015using}. Ego networks are defined by a focal node (ego) and the users to whom the ego is directly connected. High diversity score in the ego network implies that the individual splits her time evenly among her social ties.  High diversity in the network of trips among locations means that the individual distributes her number of trips evenly among her visited urban locations. The radius of gyration,  in turn,  defines the radius of the circle within which she is more likely to be found,  it is centered in all the visited locations of $i$ and weighted by the number of mobile phone records in each location \cite{gonzalez2008understanding}. For an urban science perspective,  we investigate the cell towers' residual activity as defined by \cite{toole2012inferring} to determine whether users that belong in the same cluster tent to aggregate in a specific area of the city. Residual activity can be interpreted as the amount of mobile phone activity in a region relative to the expected mobile phone activity in the whole city. Finally,  to assess the mobility behavioral pattern,  we analyze the portion of explorers and returners among the users \cite{pappalardo2015returners}. Returners are the user that limit much of their mobility to a few locations,  in contrast,  the explorers have a tendency to wander between a larger number of different locations.

\section*{Discussion}

\begin{figure}[!ht]
\centering
\includegraphics[width=\linewidth]{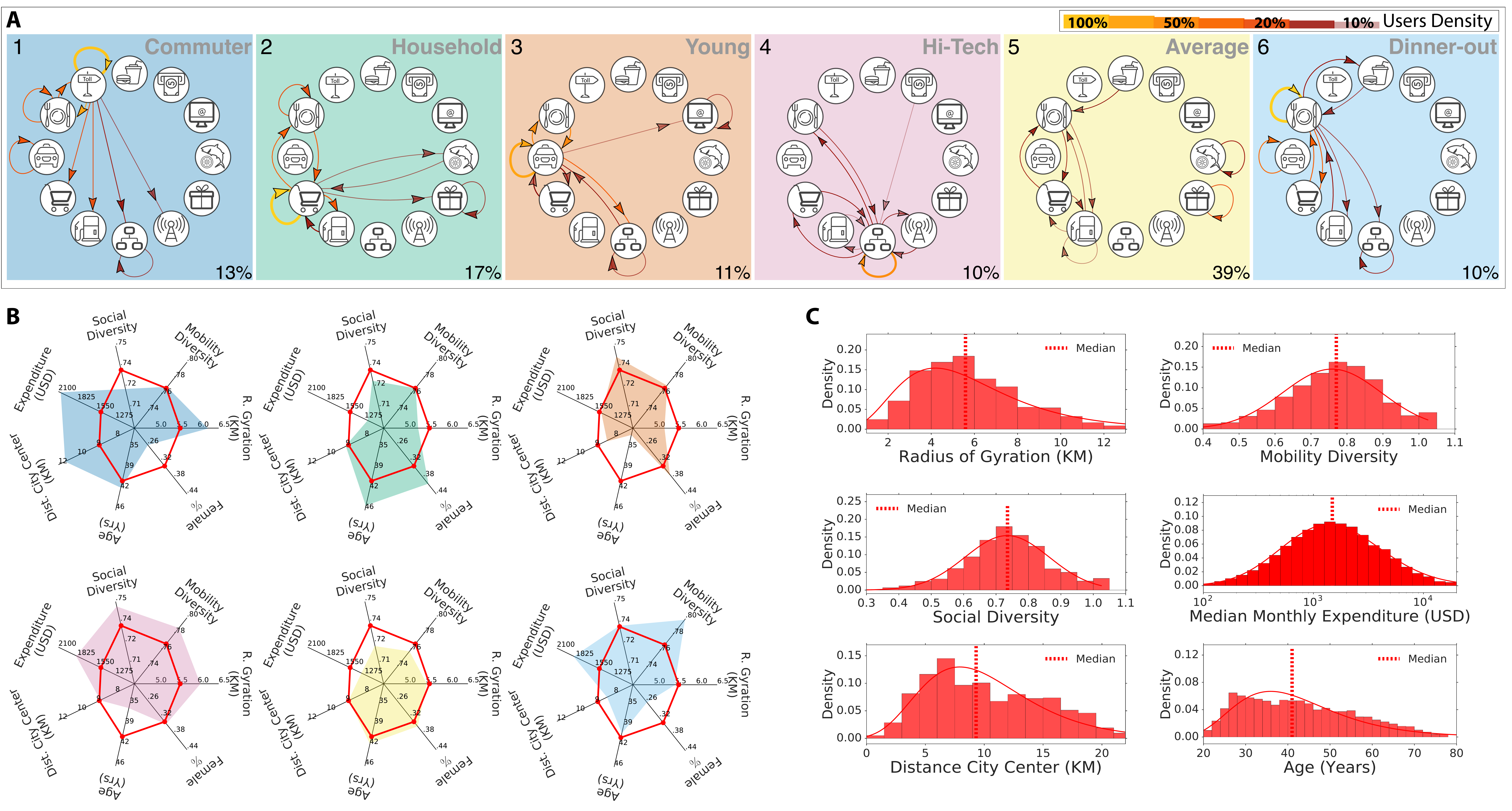}
\caption{\textbf{Identified life styles I (A)} Groups based on their spending habits. We show the top 10 most frequent spending sequences of the users in each group,  representing more than $30\%$ of users' shopping routines. The percentage of the total users in each group is shown in the bottom-right corner. \textbf{(B)} Comparison of the median of the socio-demographic variables within each group respect to the median of all users in red. (The color of the radar plot identify the spending habits in figure 4A).  \textbf{(C)} Distribution of individual characteristics among users: gender,  radius of gyration,  mobility diversity,  social diversity,  median expenditure by month,  average distance traveled from center of residence zip code to the city center and age (see Supplementary Figure  11-16, 21 for further information). The icons used in this figure are work of Azaze11o/Shutterstock.com.}
\label{fig:fig4}
\end{figure}

Five of the six clusters detected depict a particular life style on how individuals spend their money,  move and contact other individuals. One transaction type is at the core of the spending activities in each group,  and $90\%$ of the users within the cluster have it repeated as a sequence (or significant word,  represented by yellow loop in Fig.\ref{fig:fig4}A). This transaction also appears in more than $45\%$ as starting or ending transaction of the sequences of other types of transactions within the group (Fig. \ref{fig:fig4}A). The users clustered by using our approach have relatively high Shannon entropy in their transactions and a Sequitur compression ratio of 1.5 or larger (Supplementary Figure 10). Cluster 5 aggregates the uncategorized users. In particular,  the users that belong in this cluster have less than 5 significant sequences and less variation in their expenditure types (Supplementary Figure 7,  9).  

\vskip5pt

\begin{figure}[!htb]
\centering
\includegraphics[width=.97\linewidth]{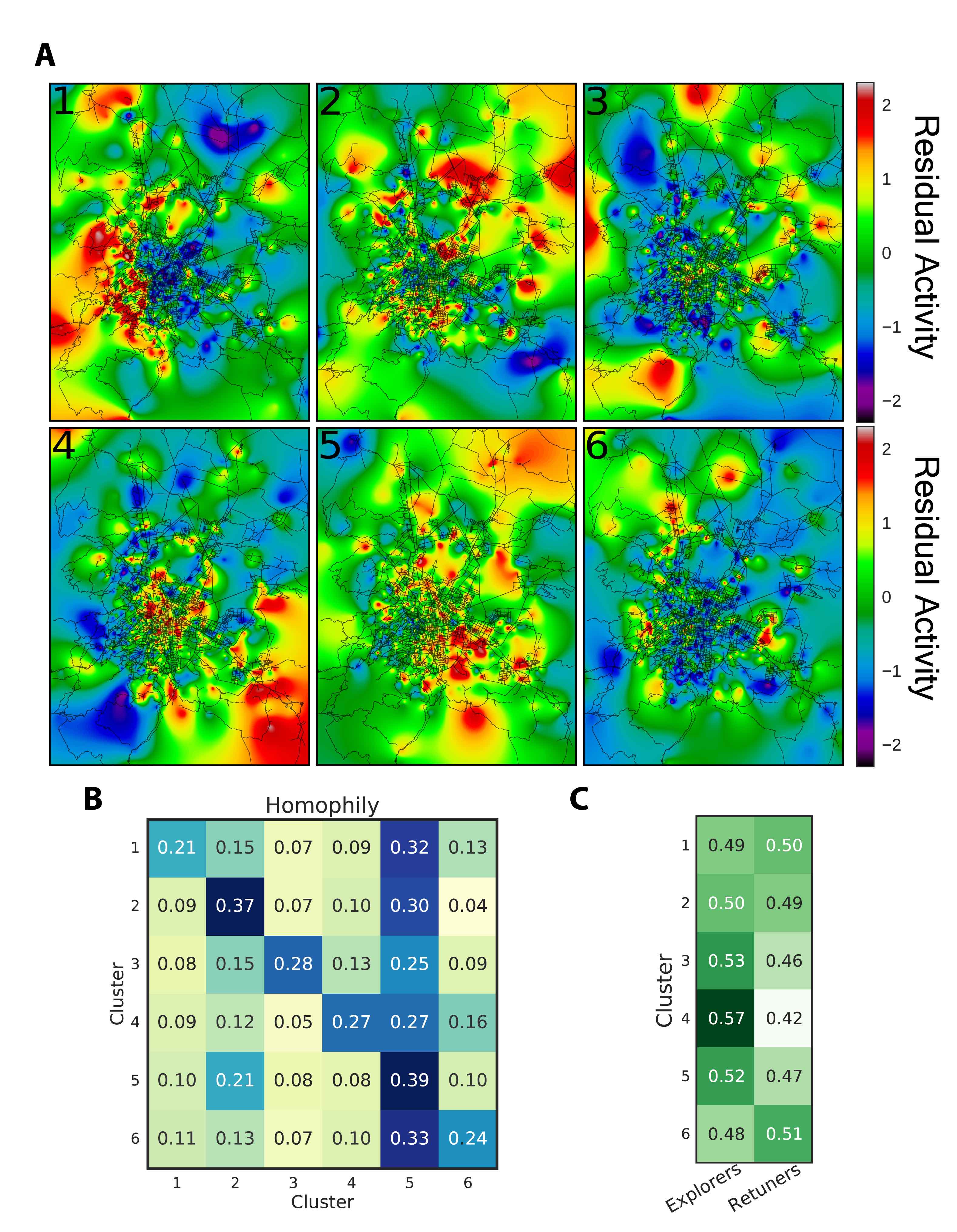}
\caption{\textbf{ Identified life styles II (A)} Cell towers residual activity by clusters. \textbf{(B)} Clusters' homophily. As expected each user tents to contact the users that belong to the same clusters or the cluster 5 ``uncategorized'' that is the cluster with the highest number of users. Remarkably there is a slight preference to contact the cluster 2 of the homemakers, that represent the oldest group. \textbf{(C)} Distribution of returners and explorers across the clusters.(see Supplementary Figure  11-16, 21 for further information). Maps in this figure were created using the software QGIS and the data from OpenStreetMap.}
\label{fig:fig5}
\end{figure}

Figures \ref{fig:fig4}B and \ref{fig:fig5} show that each cluster reveals consistent relations between expenditure patterns and the age,  mobility and social networks of their members,  hinting that the method actually unravels behavioral groups in the data,  or actual life styles. Cluster 1 aggregates users whose core transaction is toll fees and accordingly we label them as Commuters. They live furthest from the city center,  expend the most,  travel longest distances and are majority male,  as confirmed from the analysis of the radius of gyration and the residual activity Fig. \ref{fig:fig5}A. Conversely,  users in the cluster 2 or Homemakers have grocery stores as core transaction. They represent the oldest group with the least expenditure,  mobility,  and a larger share of women. Although the social network of this cluster manifests a lower diversity,  there is a slight preference in the homophily matrix to this cluster suggesting that the few connections are clusters transversal (Fig. \ref{fig:fig5}B). Younger users are split in two groups (clusters 3 and 4) with different values in their expenditure and social and mobility diversity. Cluster 3 are labeled as Youths,  because it has the youngest individuals with taxis as their core transaction. Cluster 4 is close in age to 3,  but has computer networks and information services as a core transaction. They are labeled as Tech-users and have higher than average expenditure and have higher diversity in their social contacts and mobility networks. The residual activity (Fig. \ref{fig:fig5}A) suggests that their movements are within the city center. Moreover,  the cluster 3 and 4 are the only ones with a majority of explorers within their users,  supporting the lifestyle fingerprint (Fig. \ref{fig:fig5}C). Finally,  the cluster 6,  labeled as Diners,  aggregates middle age users that have restaurants as their core transaction with high mobility diversity,  higher expenditures (see Supplementary Figure 11-16,  21 for further information).

\vskip5pt

We compare the detected groups with the ones extracted via the patients' stratification technique to analyze health records \cite{roque2011using}. Instead of applying the Sequitur algorithm to assess the likelihood of a given sequence of codes,  we compute,  for each users' code,  the TF-IDF frequency measure \cite{robertson1976relevance},  which rewards high code frequency in the individual records and penalizes high prevalence across the all users' history. The similarity matrix among users is based on cosine similarity in the space of the code frequency TF-IDF. The clusters extracted via this method (Supplementary Figure 17) do not have socio-demographic similarities and the characteristics of the members within each group average similarly to the population. Moreover,  the TF-IDF does not disentangle the Zipf-distribution  (Supplementary Figure 17c),  meaning each cluster keeps the same overall transactions' frequency. 

\vskip5pt

Furthermore we compare our clusters with the LDA \cite{blei2003latent, krestel2009latent}. This method first identifies five topics represented by an ensemble of MCCs. Each user is identified by a vector $v_i$ weighting the mixture of those five topics. We compute the uses' similarity matrix using Jensen-Shannon divergence \cite{lin1991divergence} among $v_i$. Finally,  we perform the Louvain algorithm over the matrix. Four of the seven identified clusters (1, 2, 3, 7),  in Supplementary Figure 18,  are similar to our clusters (1, 2, 3, 6).  Furthermore the LDA is able to untangle similar variance from the  Zipf distribution (Supplementary Figure 18C) compared with our method (Supplementary Figure 13B). 

\vskip5pt

Respect to the above-mentioned methods (TD-IDF,  LDA) our approach deconstructs the Zipf distribution into constituents' behavior (see Supplementary Figure 13B). The resulting clusters of the latter are comparable with our method. Furthermore,  our framework is able to capture the routines of each cluster as ordered sequence of transaction,  this temporal information is lost using the above-mentioned approaches. These tests stress the effectiveness of our method.

\vskip5pt

Finally,  we apply our framework to another minor city of Mexico: Puebla (Supplementary Figure 19-21). As already showed by \cite{sobolevsky2016cities} different cites manifest a general behavior in the term of spending patterns,  maintaining some unique characteristics. In Puebla we detect 6 clusters,  four of them share similar routines and attributes to the main city (Mexico City clusters (2, 3, 5, 6)). Comparing the Median Absolute Deviation (MAD) of each cluster is possible to assess the diversity of every socio-demographic attributes Supplementary Figure 21. In particular,  the routines of Commuters' clusters are identifiable in both of the cities with some difference in the mobility attributes. Finally,  in Puebla the Youth cluster is replaced with one with different core transactions in Miscellaneous Food store and insurance instead of taxi and restaurants. This results stresses how our framework can capture cities differences in terms of spending patterns,  providing a tool to enrich the urban activity models.

\vskip5pt

Taken together,  we present a method to detect behavioral groups in chronologically labeled data. It could be applied also to similar datasets with Zipf-like distributions,  such as disease codes in patients' visits \cite{roque2011using, hidalgo2009dynamic} or law-breaking codes in police databases \cite{schuerman1986community}. Given the ubiquitous nature of the CCRs transactions distribution by type \cite{sobolevsky2016cities},  similar groups could be detected and compared among cities worldwide. Analogous to the price index that uses online information to improve survey-based approaches to measure inflation \cite{cavallo2016scraped},  the meaningful information of groups extracted from CCRs data can be used to compare consumers worldwide \cite{mervis2012agencies}. Interesting avenues for the application of this method are policy evaluation of macroeconomic events such as inflation and employment and their effects on the spending habits of the various groups \cite{vaitla2017big}.

\section*{Methods}

\subsection*{Credit Card data sets}

Credit card data sets, also referred  as Credit Card Records CCRs, used in this study consists of 10 weeks of records starting from the 1 week of May 2015 of all the credit card user of a particular bank across each subject city. Each individual CCR consists of a hashed user identification string, the timestamp of the transaction the associated expenditure labeled with transaction type via a Merchant Category Code (MCC)  \cite{solutions2004merchant} and the transaction's amount. For each user the dataset contains age,  gender, and residential zipcode of the users  (Supplementary Figure 1A-C).The purchase entries are aggregated by user and are temporal ordered respect each day.

\subsection*{Mobile phone data sets}
Mobile phone data sets, also referred to as Call Detail Records CDRs, used in this study consist of 6 months of records starting form March 2015 of all mobile phone users of a particular carrier across each subject city. Each individual CDR consists of a hashed user identification string, a timestamp and the location of the activity. The spatial granularity of the data varies between cell tower level. 

\subsection*{Census Data}
The census data used in this work were download from the Instituto Nacional de Estad\'istica Geograf\'ia e Inform\'atica,  M\'exico (http://www.inegi.org.mx/ last checked 13/Jun/2018). In particular: the data regarding the population distribution among the districts are from ``Source: INEGI, Intercensal Survey 2015''; the data on the district income are from ``Source: INEGI, National Survey of Occupation and Employment (ENOE). Population aged 15 years and older.''

\subsection*{Data availability}
For contractual and privacy reasons,  the raw data cannot be available. Upon request, the authors can provide the data of the matrix of user similarity along with the appropriate documentation for replication.

\section*{Acknowledgements}

This work was supported by Gates Foundation (grant OPP1141325) and United Nations Foundation (grant UNF-15-738). We acknowledge Rebecca Furst-Nichols,  and Jake Kendall on the planning of the study. We also thank Edward Barbour,  Philip Chodrow and Balazs Lengyel for the helpful discussions. Views and conclusions in this document are those of the authors and should not be interpreted as representing the policies,  either expressed or implied,  of the sponsors. Riccardo Di Clemente as Newton International Fellow of the Royal Society acknowledges support from The Royal Society,  The British Academy and the Academy of Medical Sciences (Newton International Fellowship,  NF170505). The icons used in this paper are work of Azaze11o/Shutterstock.com.

\section*{Author contributions statement}

R.D.C analyzed the data,  performed the research, R.D.C created the maps, S.X. developed and tested the machine learning algorithm; R.D.C.,  M.T.,  M.L.O.,  B.V. and M.C.G. planned the study; R.D.C. and M.C.G. design the study and wrote the paper.; M.C.G. coordinated the study. All authors gave final approval for publication. 

\section*{Competing Interests}

The Authors declare no competing interests.
The corresponding author is responsible for submitting a \href{http://www.nature.com/srep/policies/index.html#competing}{competing interests statement} on behalf of all authors of the paper.

\section*{Figures}

\clearpage

\section*{Supporting information}

\beginsupplement

\section*{Figures}

\begin{figure}[!htb]
\centering
\includegraphics[width=\linewidth]{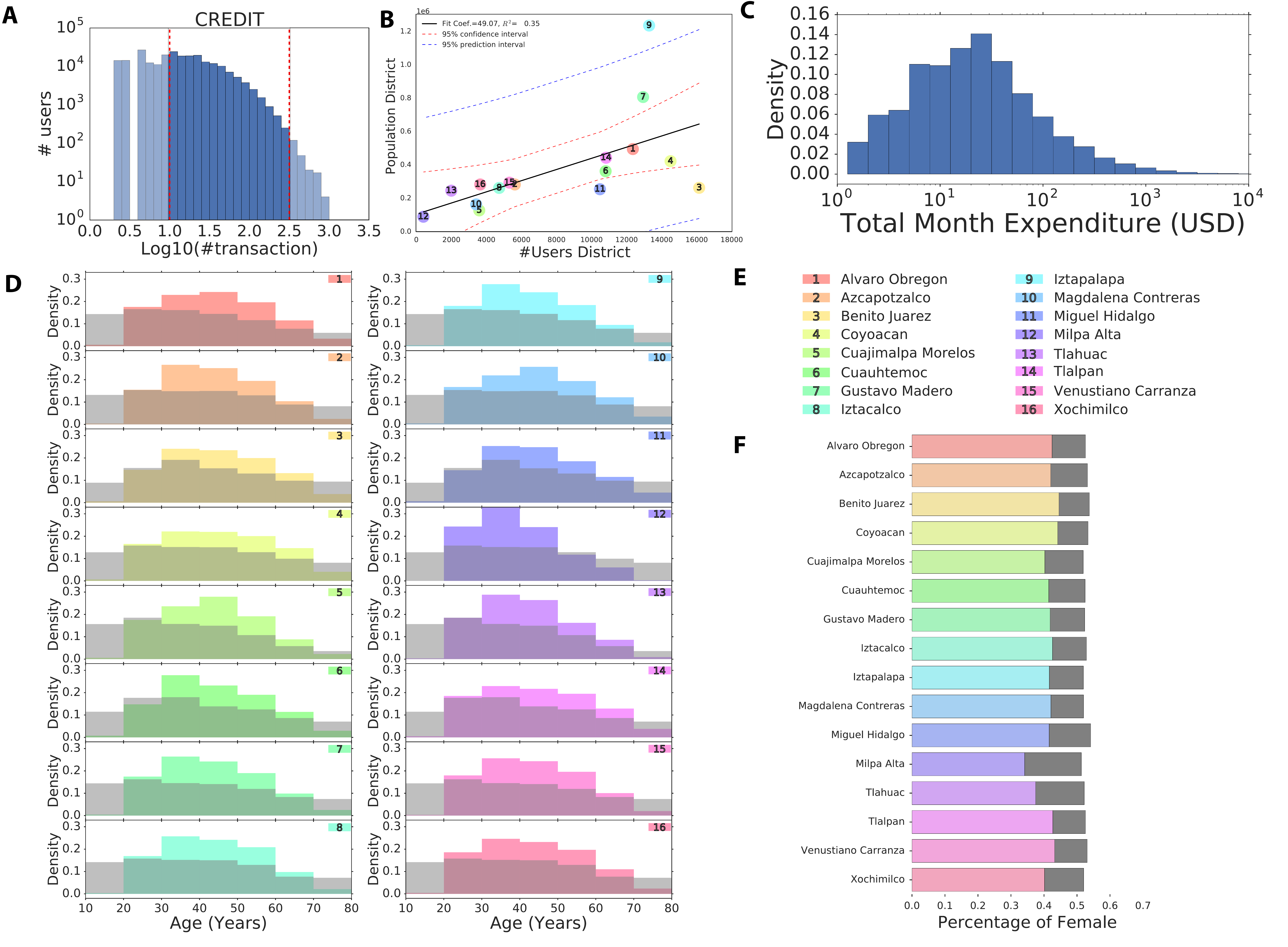}
\caption{\textbf{(A)} Histogram of the user transaction number in the CCRs, the 150,000 users selected for the analysis are those with more than 10 transactions and less than 300. \textbf{(B)} Relation between the district population ("Source: INEGI, Intercensal Survey 2015'') and the number of user in our datasets, using the same color map as Fig. 1 in the paper (each color represents a city district Fig. 1A main text, the districts legend in Fig. S1s). \textbf{(C)} Distribution of overall users' monthly expenditure in USD. \textbf{(D)} Comparison between district users of CCR and district population from Census Data in gray ("Source: INEGI, Intercensal Survey 2015''). \textbf{(E)} Mexico City Color legend district for Fig. 1A of the main text and figures S1b-S1d-S1F. \textbf{(F)} Female percentage usage of credit card (by district) CCRs, in comparison with the female percentage from census in gray ("Source: INEGI, Intercensal Survey 2015'').}
\label{fig:Sfig1}
\end{figure}

\begin{figure}[!htb]
\centering
\includegraphics[width=0.8\linewidth]{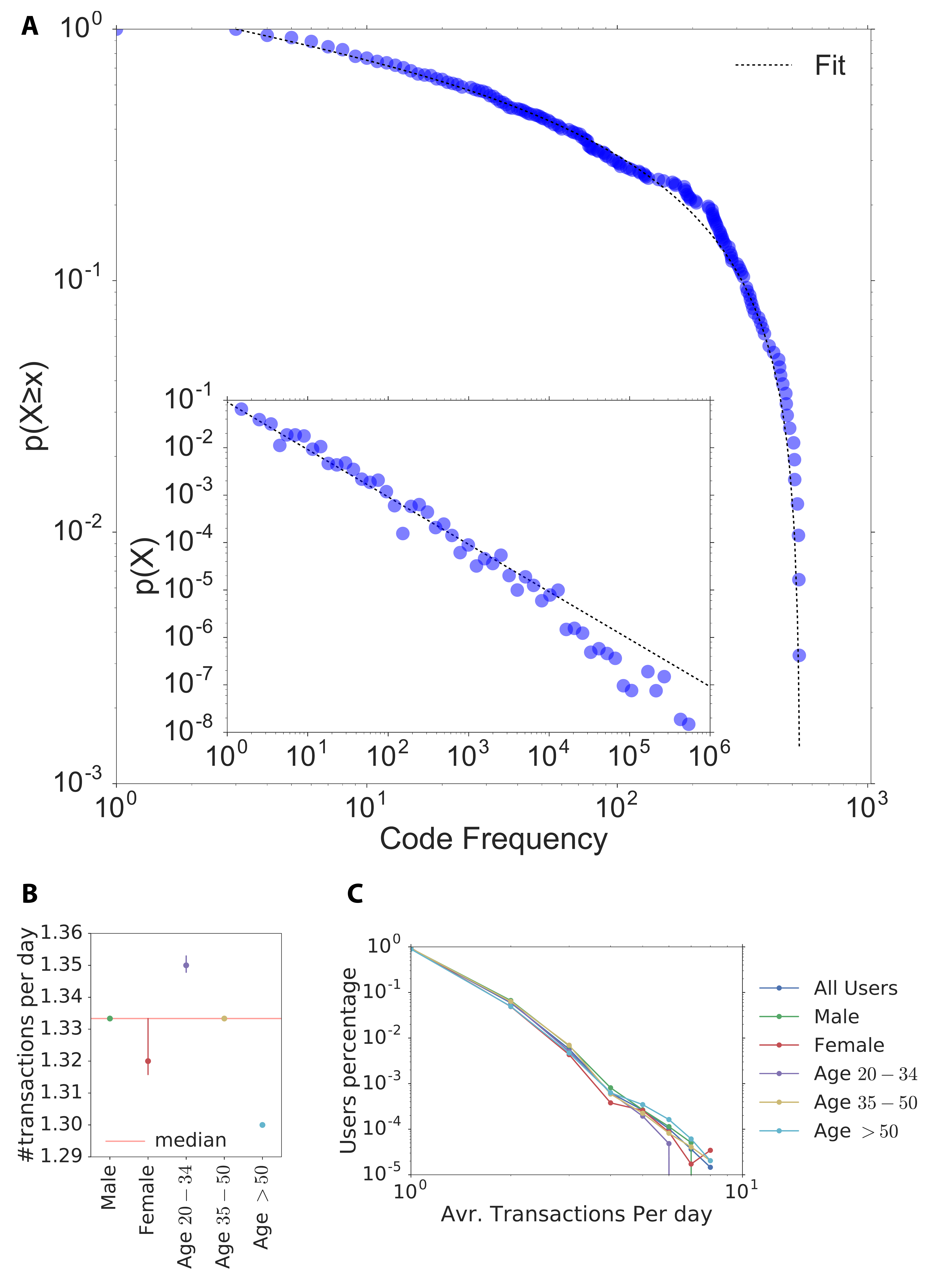}
\caption{\textbf{(A)} Complementary Cumulative Distribution and Probability Density Plot (inset) of MMCs' transaction
codes. The probability distribution of a transaction code $x$ presents Zipf's distribution $p(w_i)\propto x_i^{(-1.05)}$, with a Kolmogorov-Smirnov distance $D_{n=0.04}$ and with cutoff identified as the right-most point in the distribution before the fitted power law \cite{alstott2014powerlaw,clauset2009power}. \textbf{(B)} Median's distributions of the number of transactions per day divided by socio-demographic features; the error bars represent the confidence interval of $95\%$  \textbf{(C)} Percentage of the number of transactions per day per user divided by socio-
 demographic features.}
\label{fig:Sfig2}
\end{figure}

\begin{figure}[!htb]
\centering
\includegraphics[width=0.8\linewidth]{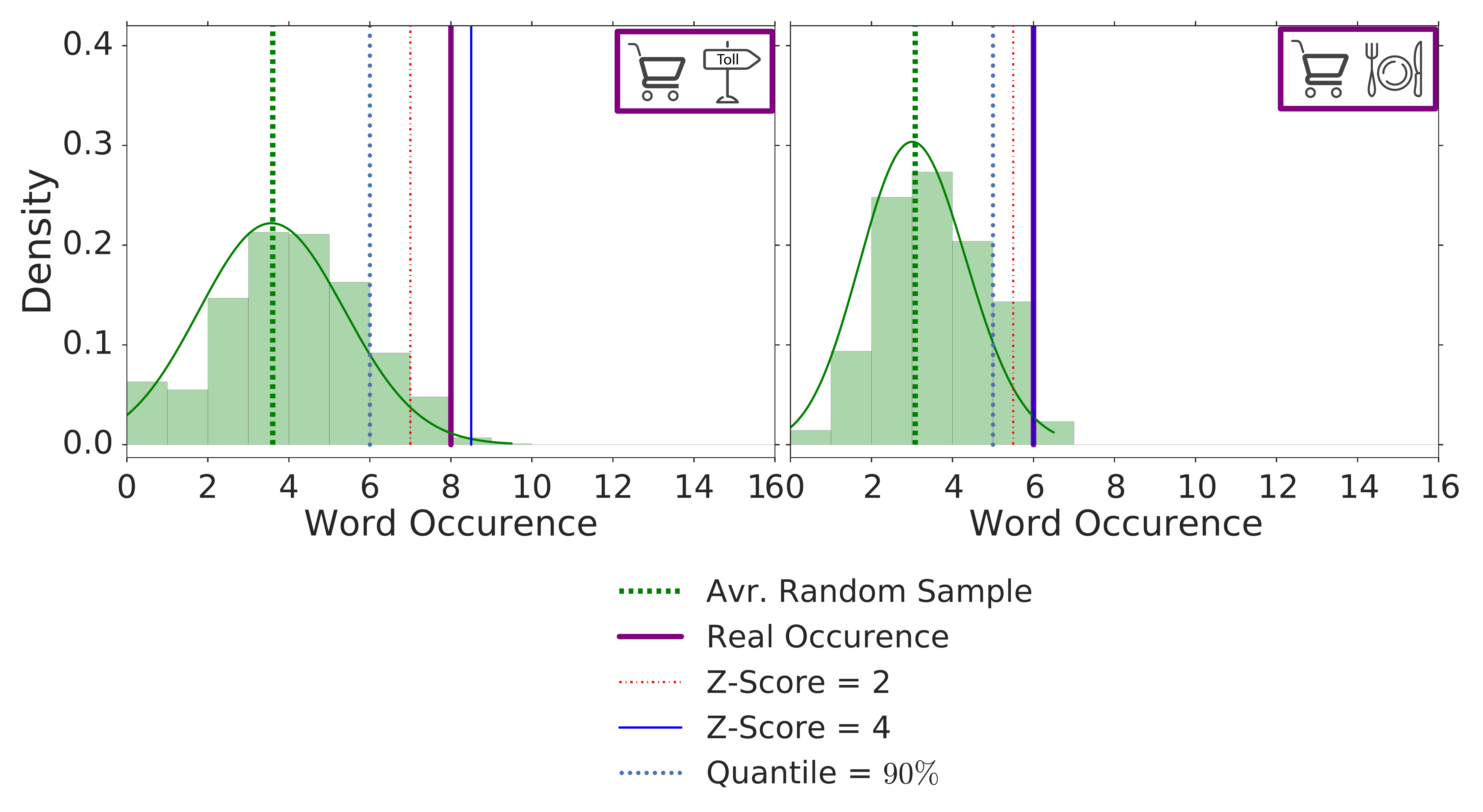}
\caption{Example of word occurrence distribution of the 1000 randomized code sequences preserving the same number of transactions per type. In both cases the real occurrence of the word showed in the purple box his higher than the average of the random sample. We can see that the z-score equal 2 computed from the sample in relation with the 90th quantile of the distribution. The icons used in this figure are work of Azaze11o/Shutterstock.com.}
\label{fig:Sfig3}
\end{figure}

\begin{figure}[!htb]
\centering
\includegraphics[width=0.6\linewidth]{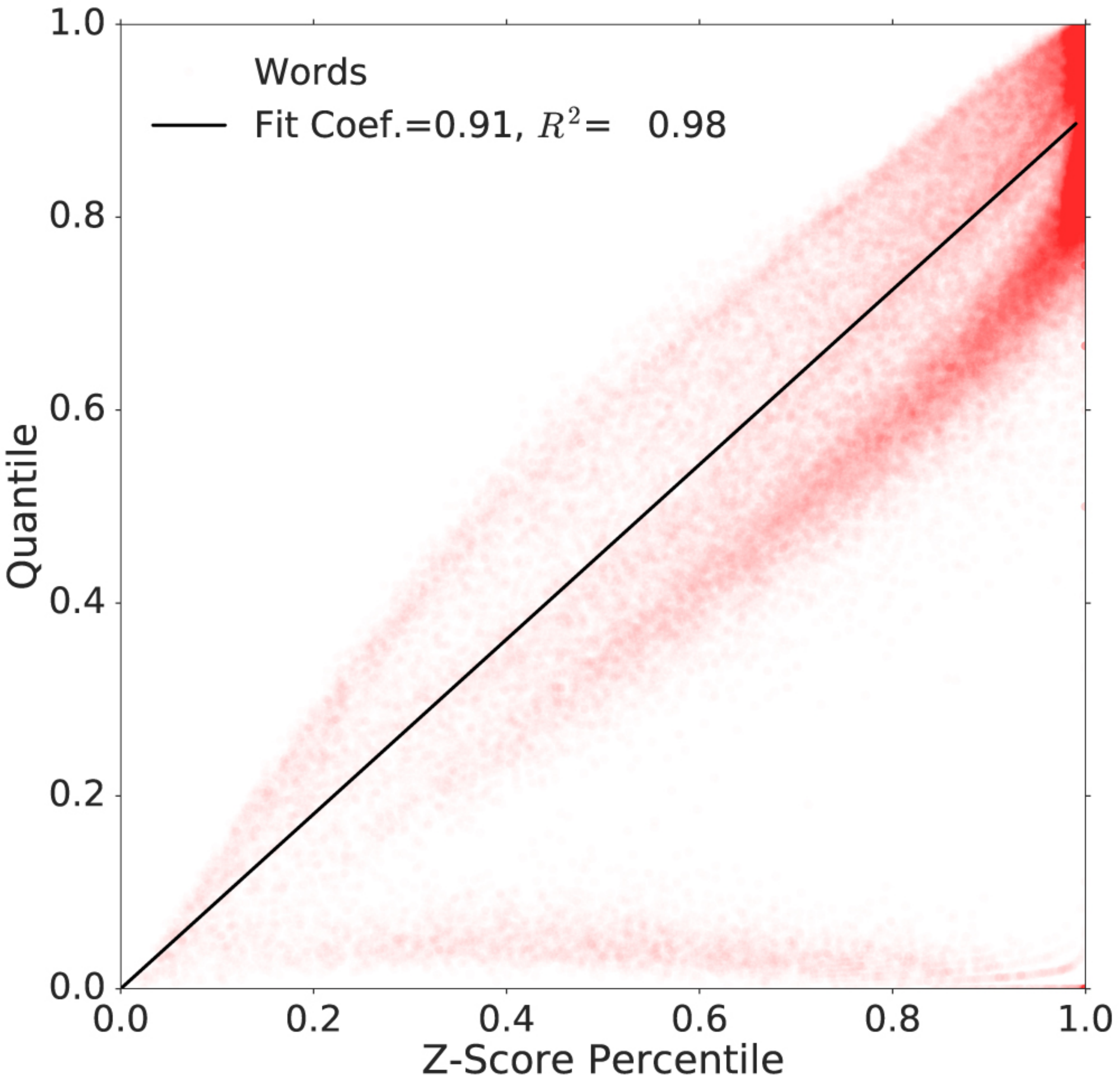}
\caption{Relation between the quantile value of the real occurrence of words respect to the randomized distribution vs. the z-score relative percentile of the words. The z-score relative percentile is highly correlated with the quantile position of the real word occurrence. We selected only the words with z-score$>2$, corresponding to the 97.73th percentile for a Gaussian distribution.}
\label{fig:Sfig4}
\end{figure}

\begin{figure}[!htb]
\centering
\includegraphics[width=0.87\linewidth]{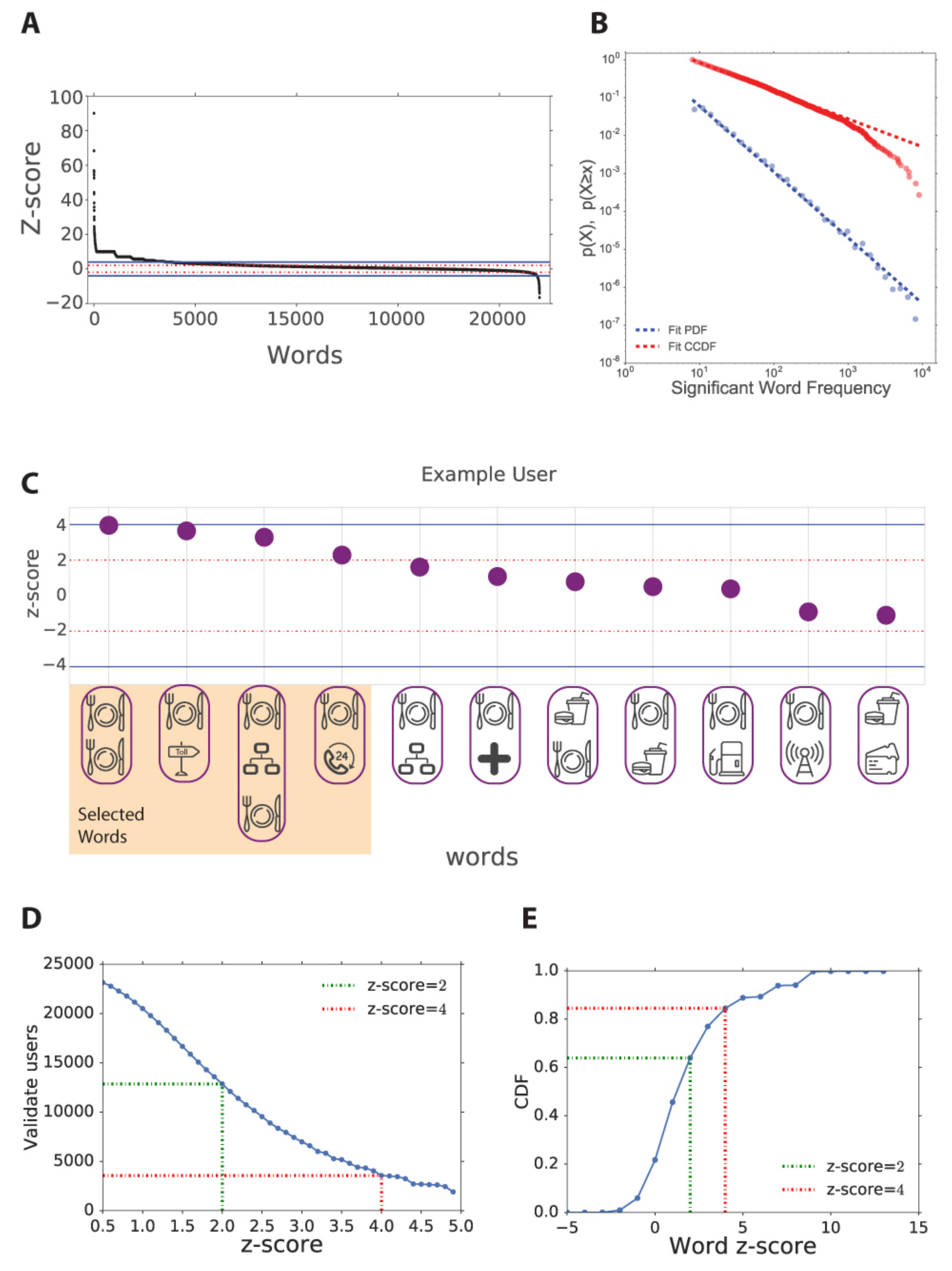}
\caption{\textbf{(A)} The z-score of each word depend on occurrence. The z-score of each word is calculated by comparing
the code sequence with $1000$ randomized code sequences for each user, preserving the number of transactions per type. \textbf{(B)} Probability Density Function plot of the occurrence of the significant words $\{W_{i}\}$ and its complementary cumulative distribution; the probability distribution words manifest a power law behavior $p(W_i)\propto x_i^{(-1.73)}$ with $x_{i}$ frequency of the $\{W_{i}\}$ and Kolmogorov-Smirnov test of $0.02$. \textbf{(C)} Z-score of a words occurrence for a sample user. We highlight in orange the users' associated set of words with a z-score greater than 2, that characterizes the user shopping routines. \textbf{(D)} Number of validate users given a z-score threshold. Selecting higher values of the z-score generate a decrease of analyzable users, while low values will impact the significance of the word selected. \textbf{(E)} CDF of the words z-scores. The icons used in this figure are work of Azaze11o/Shutterstock.com.}
\label{fig:Sfig5}
\end{figure}

\begin{figure}[!htb]
\centering
\includegraphics[width=0.99\linewidth]{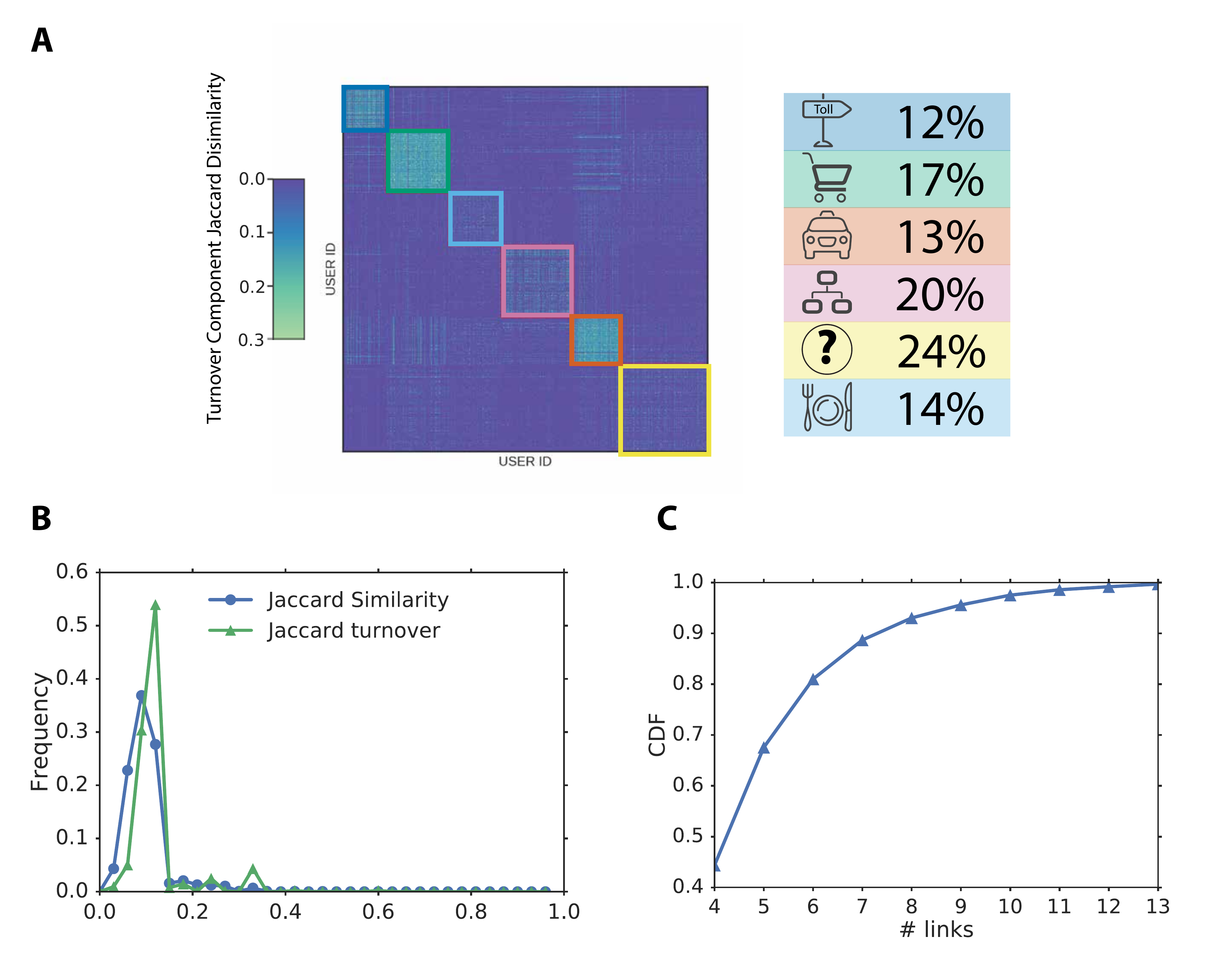}
\caption{\textbf{(A)} Clustering results using the turnover component of the Jaccard dissimilarity index \cite{baselga2012relationship}. \textbf{(B)} Distributions comparison between the Jaccard Similarity and the Jaccard turnover. \textbf{(C)} CDF of the number of links per users' network. The icons used in this figure are work of Azaze11o/Shutterstock.com.}
\label{fig:Sfig6}
\end{figure}

\begin{figure}[!htb]
\centering
\includegraphics[width=0.99\linewidth]{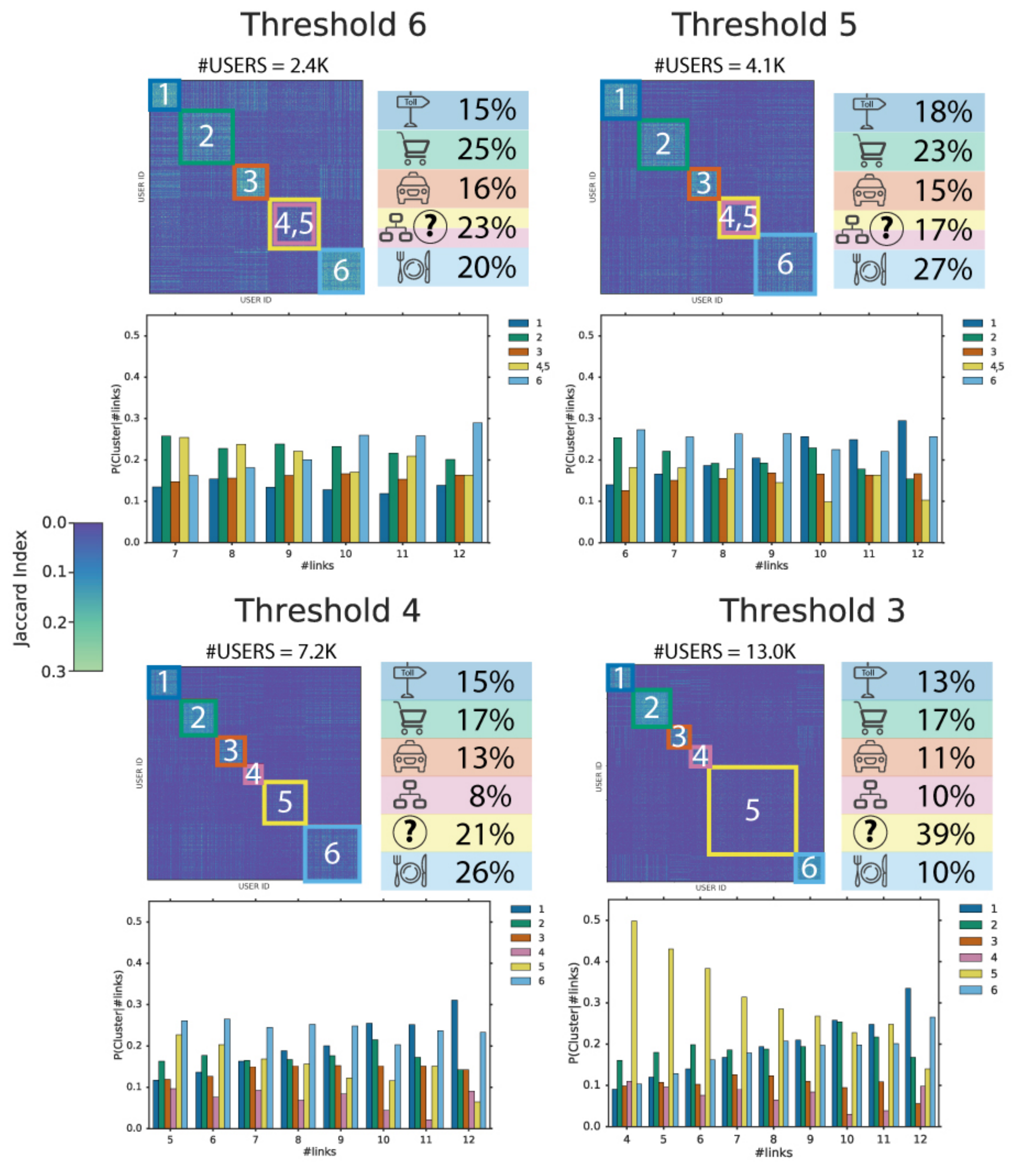}
\caption{Clustering results depending on the users' selection. For each threshold $x$ we select all the users with more than $x$ significant links. Using the Louvain Algorithm \cite{blondel2008fast,staudt2016engineering} we perform the cluster of the users' similarity matrix of the selected users at each threshold. For each threshold, we show the proportion of users that belong to each cluster, the core transaction for that cluster (as defined in the main text, Fig. 3 main text) and the conditional probability for a user to belong in a given cluster depending on its number of significant links \textbf{$P(cluster|\# links)$}. By applying a lower threshold is it possible to increase the number of users analyzed. In particular, selecting users with more than 3 significant links improves the identification of clusters 4 and 6, which were misidentified when using higher thresholds. At the same time lowering the threshold increases the number of user that we are not able to categorize effectively (user percentage of cluster 5). The icons used in this figure are work of Azaze11o/Shutterstock.com.}
\label{fig:Sfig7}
\end{figure}

\begin{figure}[!htb]
\centering
\includegraphics[width=0.99\linewidth]{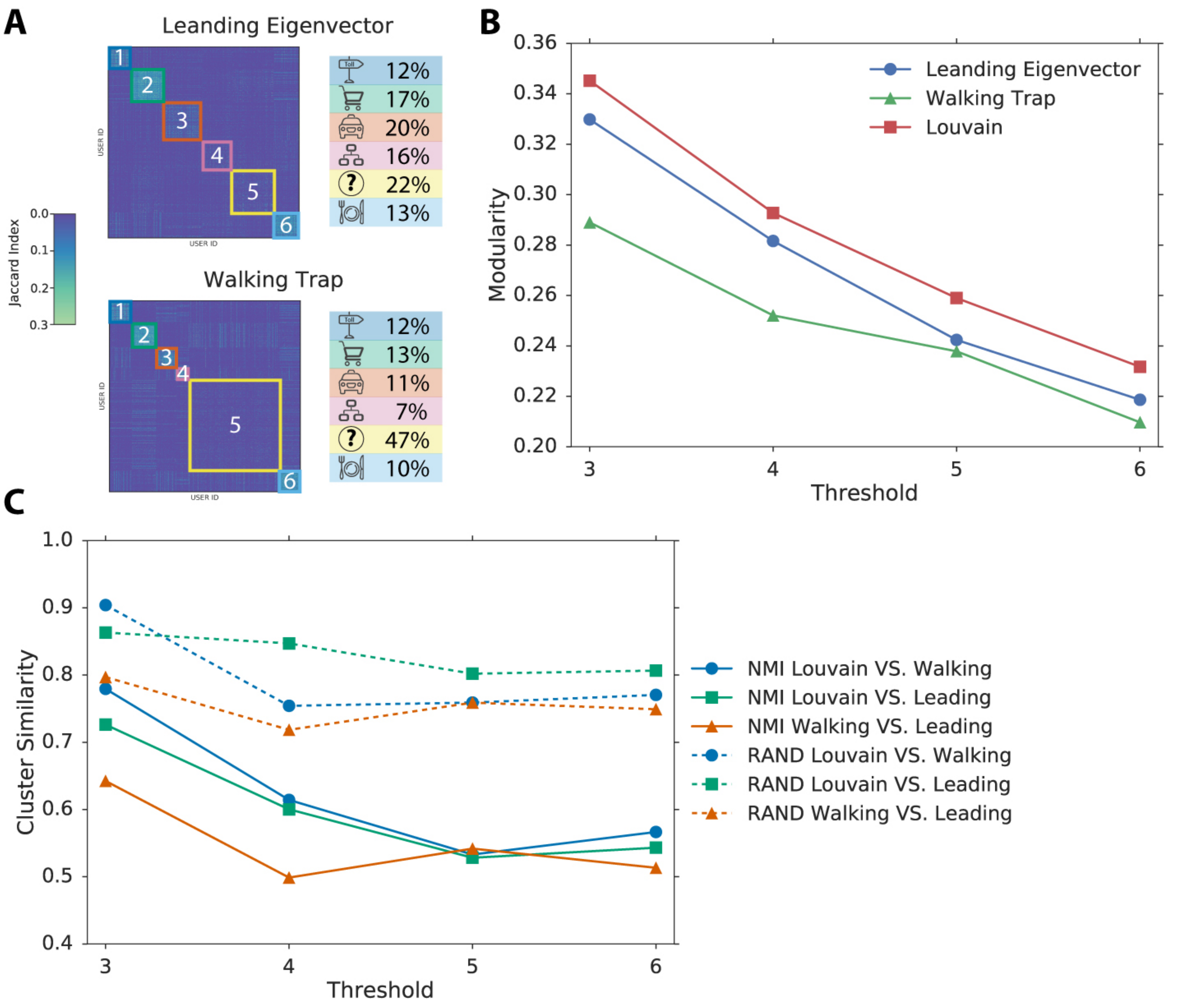}
\caption{\textbf{(A)} Cluster analysis of the users' similarity matrix for threshold 3 using the Leading Eigenvector \cite{newman2006finding} and Walking Trap \cite{pons2005computing} algorithms; both algorithms detect six different clusters as Louvain \cite{blondel2008fast} (Fig. 3 main text and Supplementary Figure 5). \textbf{(B)} Network modularity analysis depending on the three cluster algorithms proposed. We see the Louvain algorithm always performs better in terms of modularity \cite{newman2006modularity}. \textbf{(C)} Analysis of similarity between the three methods of data clustering performed, using Normalized Mutual Information (NMI) \cite{danon2005comparing,ana2003robust} and Rand \cite{rand1971objective} index. Values near one suggest a higher similarity between the cluster identified by the Louvain algorithm and the other two algorithms. The icons used in this figure are work of Azaze11o/Shutterstock.com.}
\label{fig:Sfig8}
\end{figure}

\begin{figure}[!htb]
\centering
\includegraphics[width=0.99\linewidth]{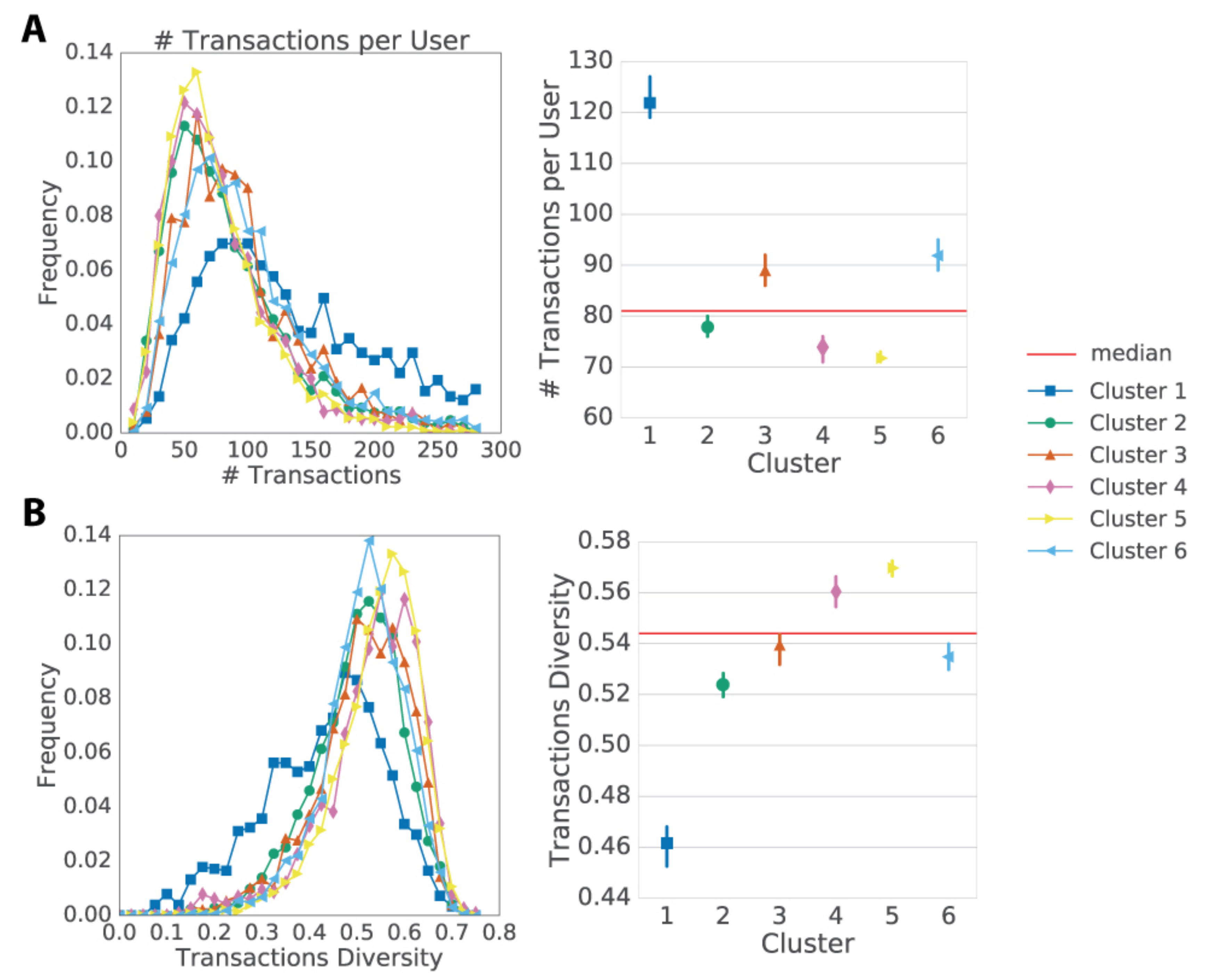}
\caption{\textbf{(A)} Distributions of the number of users' transactions per clusters (on the left), and confidence interval of $95\%$
(on the right). \textbf{(B)} Distribution of the users' transaction diversity per clusters(on the left), and confidence interval of $95\%$ (on the right). We measure the transaction diversity
$D(i)$ of a user $i$ by using the Shannon entropy of the user's transactions and dividing by the number of transactions
$N$ hence: $D(i)=[\sum_{t_{i}\in T_{i}}p(t_{i})\log p(t_{i})]/N$; with $T$ the set of user transaction. The users identified as ``commuters'' (see main text) are those with low transactions diversity and a higher frequency of transactions. Conversely the users in the cluster 5 manifest a higher transaction diversity with a low number of transactions. These two factors combined means that the identification of the users' routines in this cluster is more challenging. }
\label{fig:Sfig9}
\end{figure}

\begin{figure}[!htb]
\centering
\includegraphics[width=0.99\linewidth]{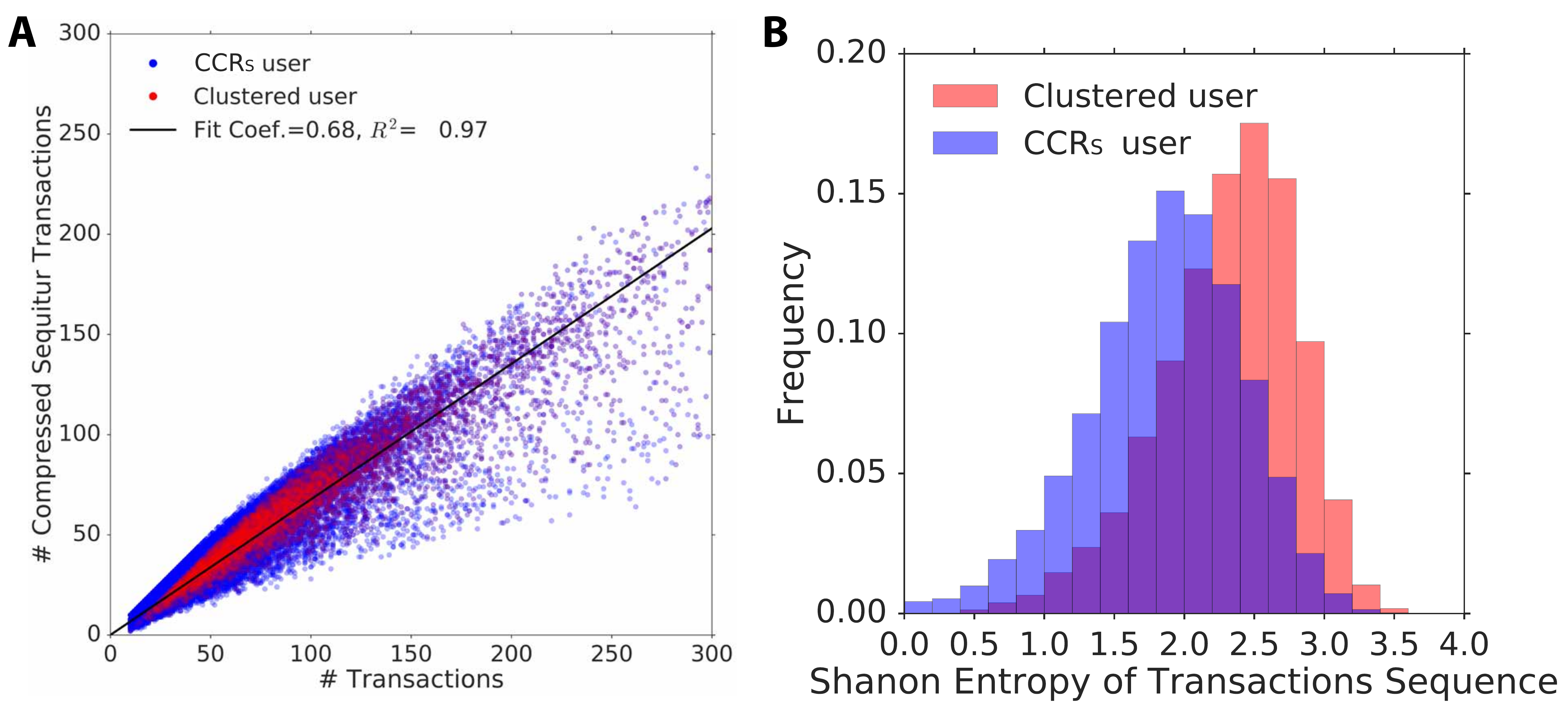}
\caption{\textbf{(A)}  Sequitur compression ratio. Ratio between the original sequence transactions length and the length of Sequitur \cite{nevill1997identifying} sequence output. The compression ratio of the clustered user is 1.50. \textbf{(B)} Shanon entropy of the transactions Sequence. We define the Shannon entropy for a user $i$ as $S(i) = [\sum_{t_{i}\in T_{i}}p(t_{i})\log p (t_{i})]$.}
\label{fig:Sfig10}
\end{figure}

\begin{figure}[!htb]
\centering
\includegraphics[width=0.99\linewidth]{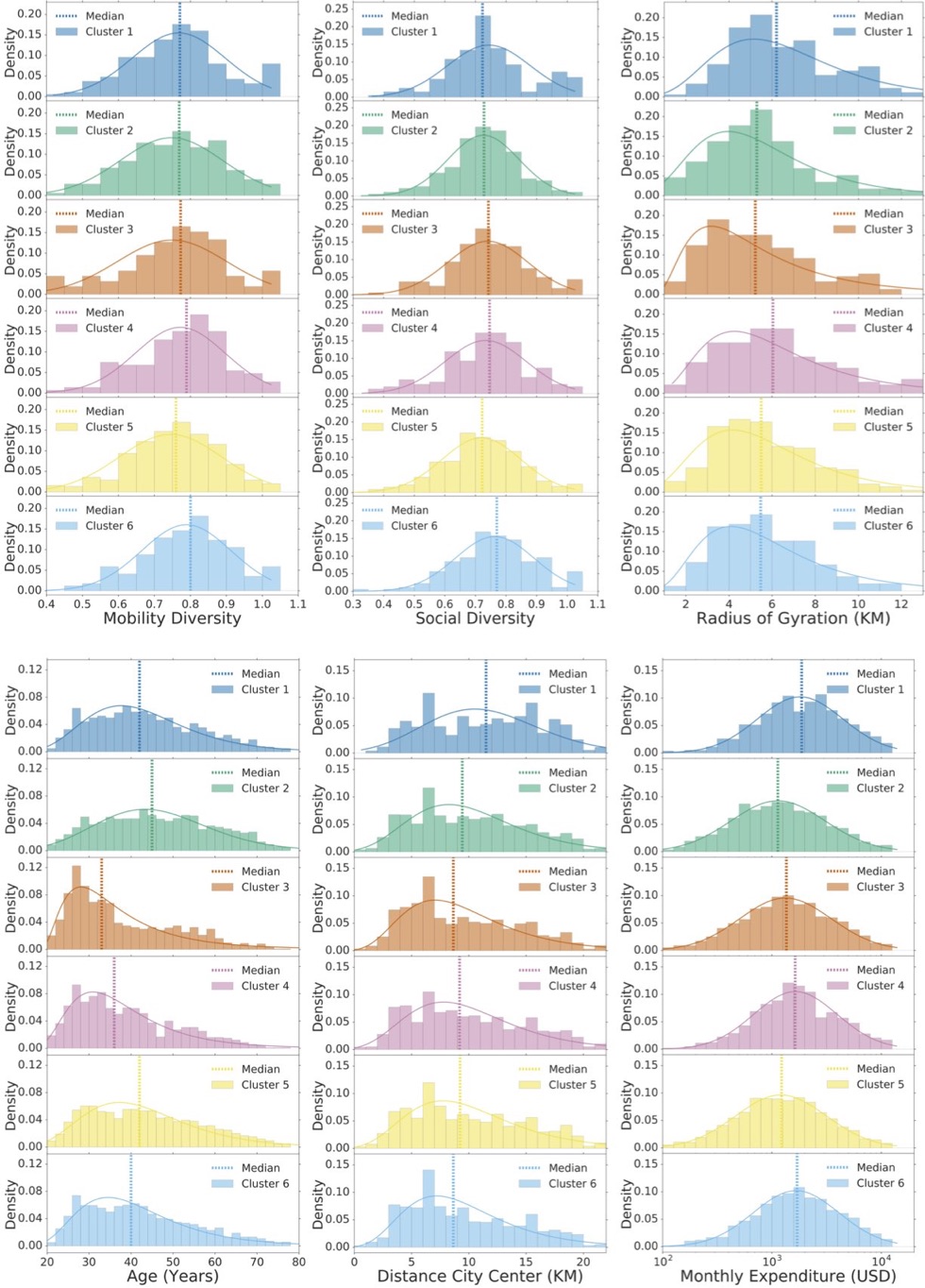}
\caption{Distributions of socio-demographic characteristics of individuals in each cluster.}
\label{fig:Sfig11}
\end{figure}

\begin{figure}[!htb]
\centering
\includegraphics[width=0.99\linewidth]{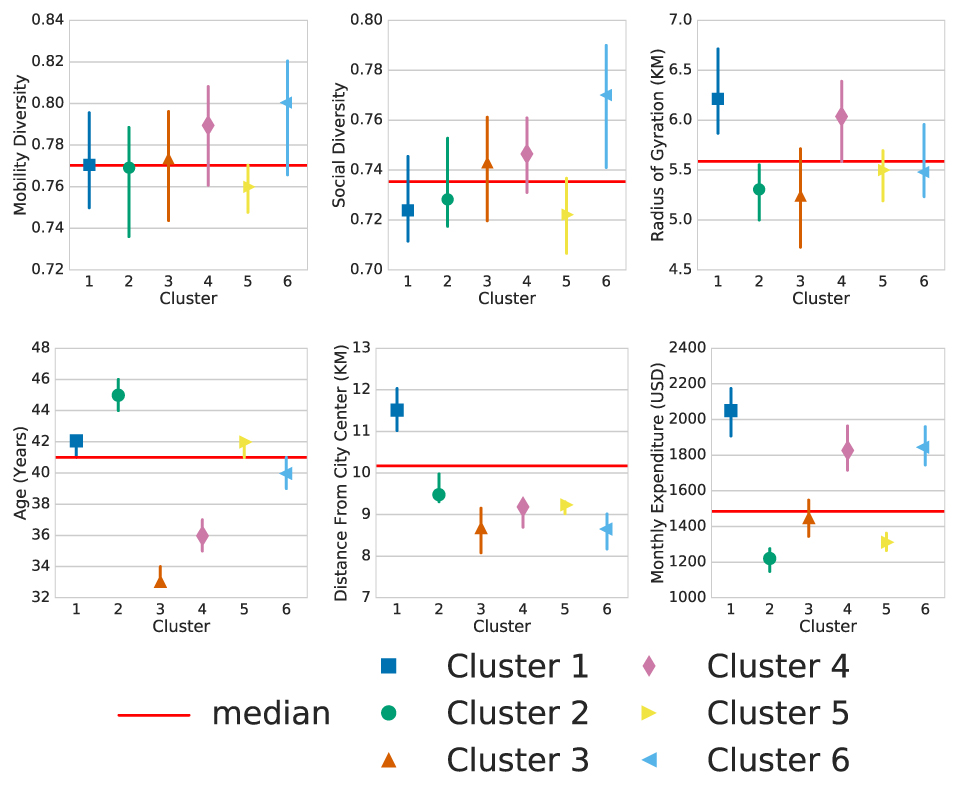}
\caption{Confidence intervals of $95\%$ of socio-demographic characteristics of individuals in each cluster detected by our framework, and the solid in red representing the median values of the all clustered users.}
\label{fig:Sfig12}
\end{figure}

\begin{figure}[!htb]
\centering
\includegraphics[width=0.99\linewidth]{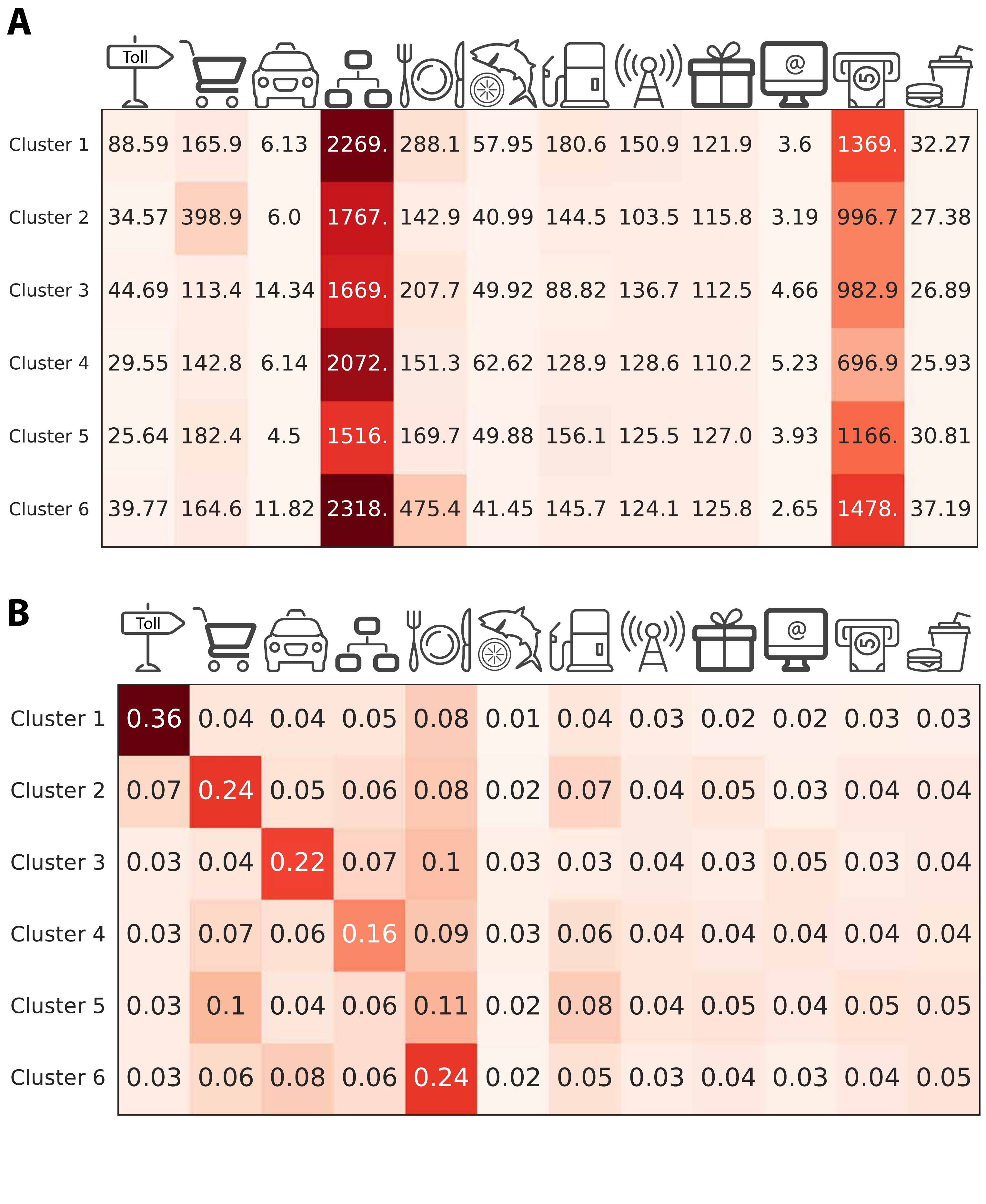}
\caption{\textbf{(A)} Median expenditure by transaction code (in USD for the 10 weeks considered). The overall clusters' expenditure are in agreement with the core transaction identified by our framework. \textbf{(B)} Frequency of transaction code for the 10 weeks considered for each fo the six cluster detected; the core transaction is a dominant feature for each of the clusters. Our method is able to extract information form a zipf like distribution uncovering behavior in shopping patterns. The icons used in this figure are work of Azaze11o/Shutterstock.com.}
\label{fig:Sfig13}
\end{figure}

\begin{figure}[!htb]
\centering
\includegraphics[width=0.6\linewidth]{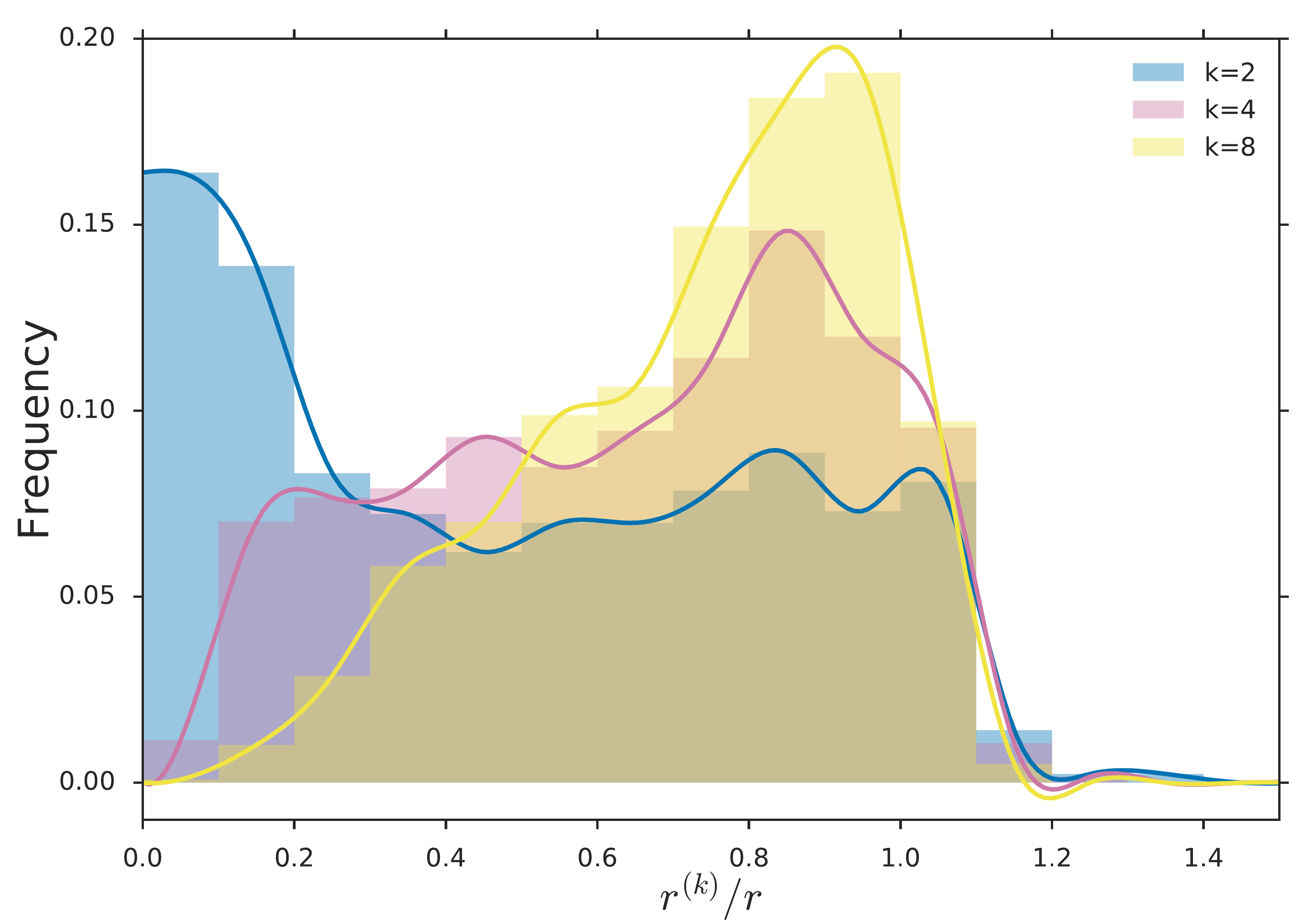}
\caption{Returners and Explorer \cite{pappalardo2015returners} analysis all the users.}
\label{fig:Sfig14}
\end{figure}

\begin{figure}[!htb]
\centering
\includegraphics[width=0.99\linewidth]{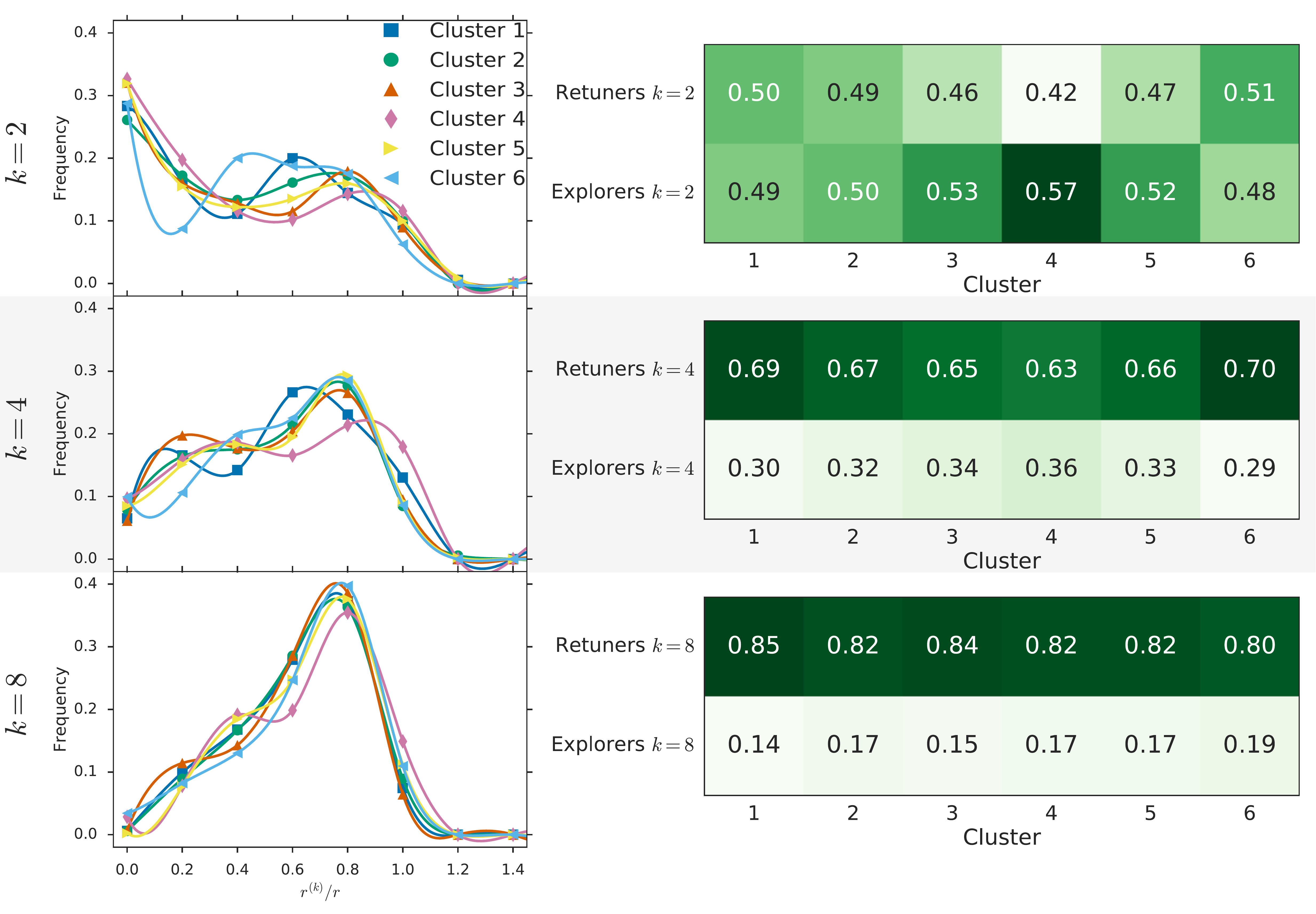}
\caption{Returners and Explorer \cite{pappalardo2015returners} analysis by clusters.}
\label{fig:Sfig15}
\end{figure}

\begin{figure}[!htb]
\centering
\includegraphics[width=0.99\linewidth]{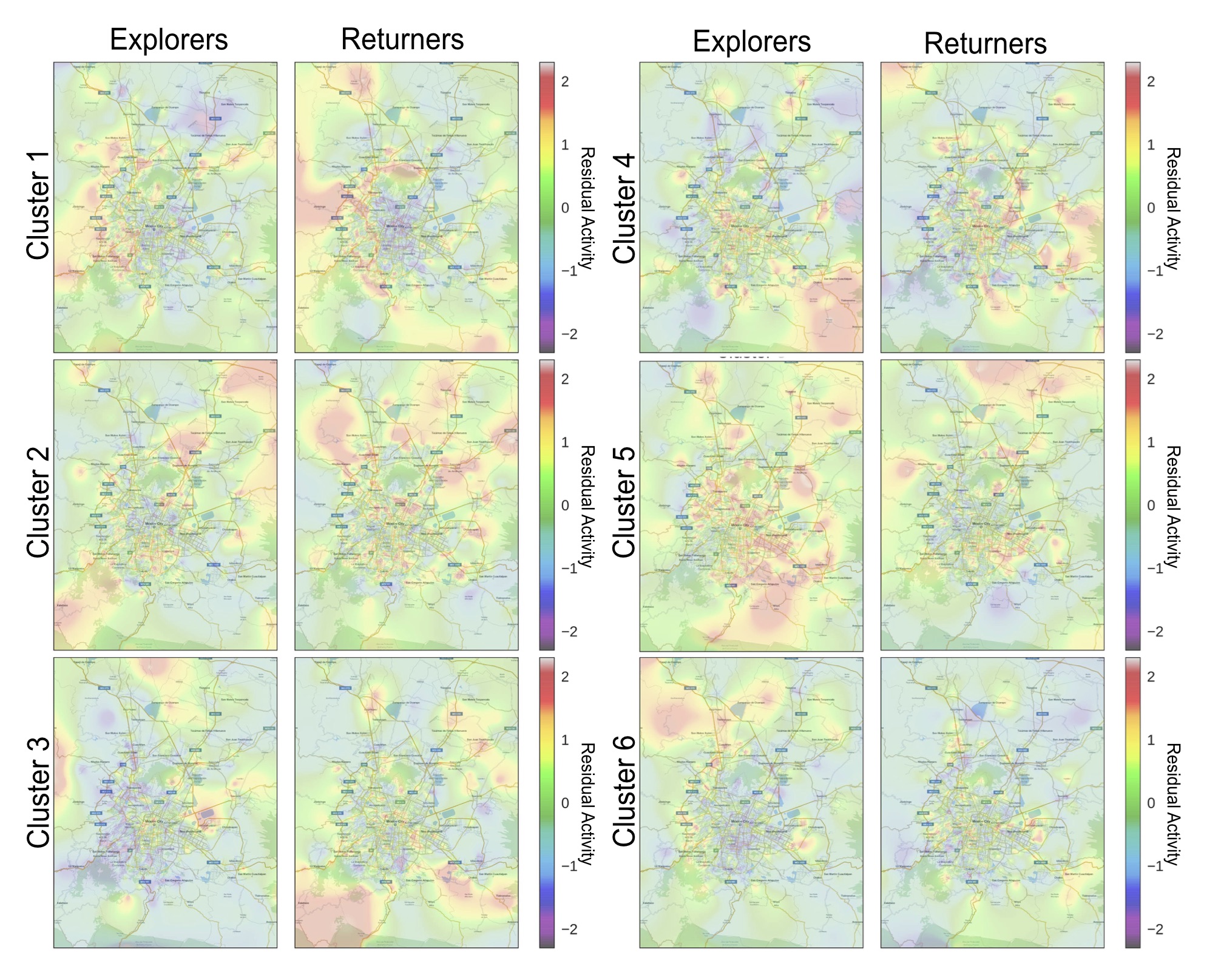}
\caption{Returners and Explorer \cite{pappalardo2015returners} cell tower residual activity \cite{toole2012inferring}. The maps were created using the software Mathematica}
\label{fig:Sfig16}
\end{figure}

\begin{figure}[!htb]
\centering
\includegraphics[width=\linewidth]{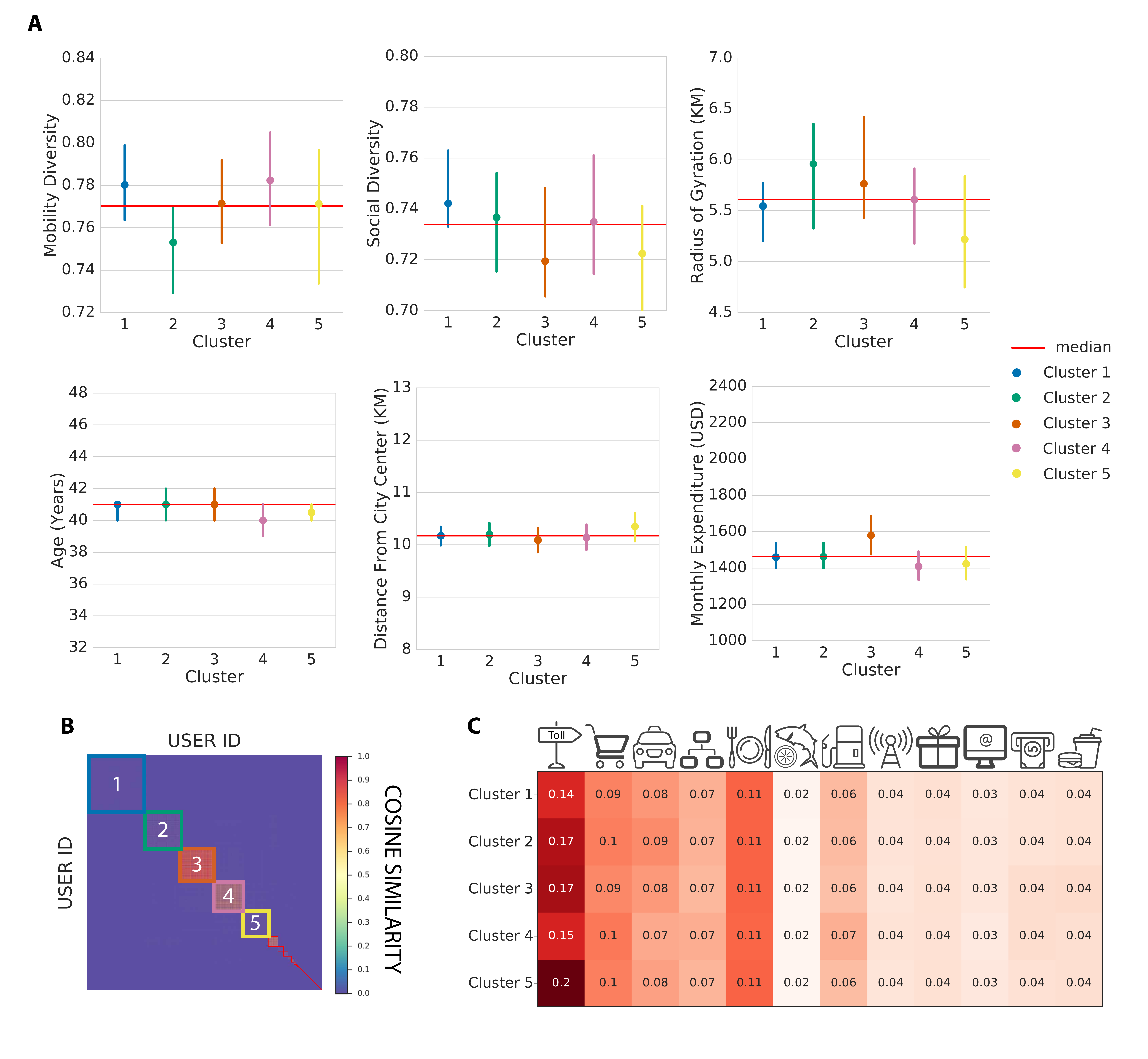}
\caption{Cluster analysis of the selected user using the TF-IDF \cite{roque2011using,hidalgo2009dynamic} algorithm to compare the users transaction code sequence. \textbf{(A)} Confidence interval of $95\%$ of socio-demographic characteristics of the five clusters detected. The TF-IDF is not able to capture any hidden information form the zipf like distribution of the credit card transaction. According with our socio-demographic metrics each of the 5 cluster detected is only a random sample of users and do not show any particular behavior. \textbf{(B)} Cluster analysis of the 13.0K selected users. The Louvain algorithm has been performed over the users' cosine similarity matrix of the TF-IDF, with a threshold at 0.6 of cosine similarity  \cite{roque2011using}. The clusters 3,4 show and high users' similarity in the TF-IDF without showing any meaningful socio-demographic relation. \textbf{(C)} Transactions frequency for each clusters show the same zipf like distributions outlining that standard methods are not suitable to extract information form a zipf like distribution as our framework (see Supplementary Figure 13B). The icons used in this figure are work of Azaze11o/Shutterstock.com.}
\label{fig:Sfig17}
\end{figure}

\begin{figure}[!htb]
\centering
\includegraphics[width=\linewidth]{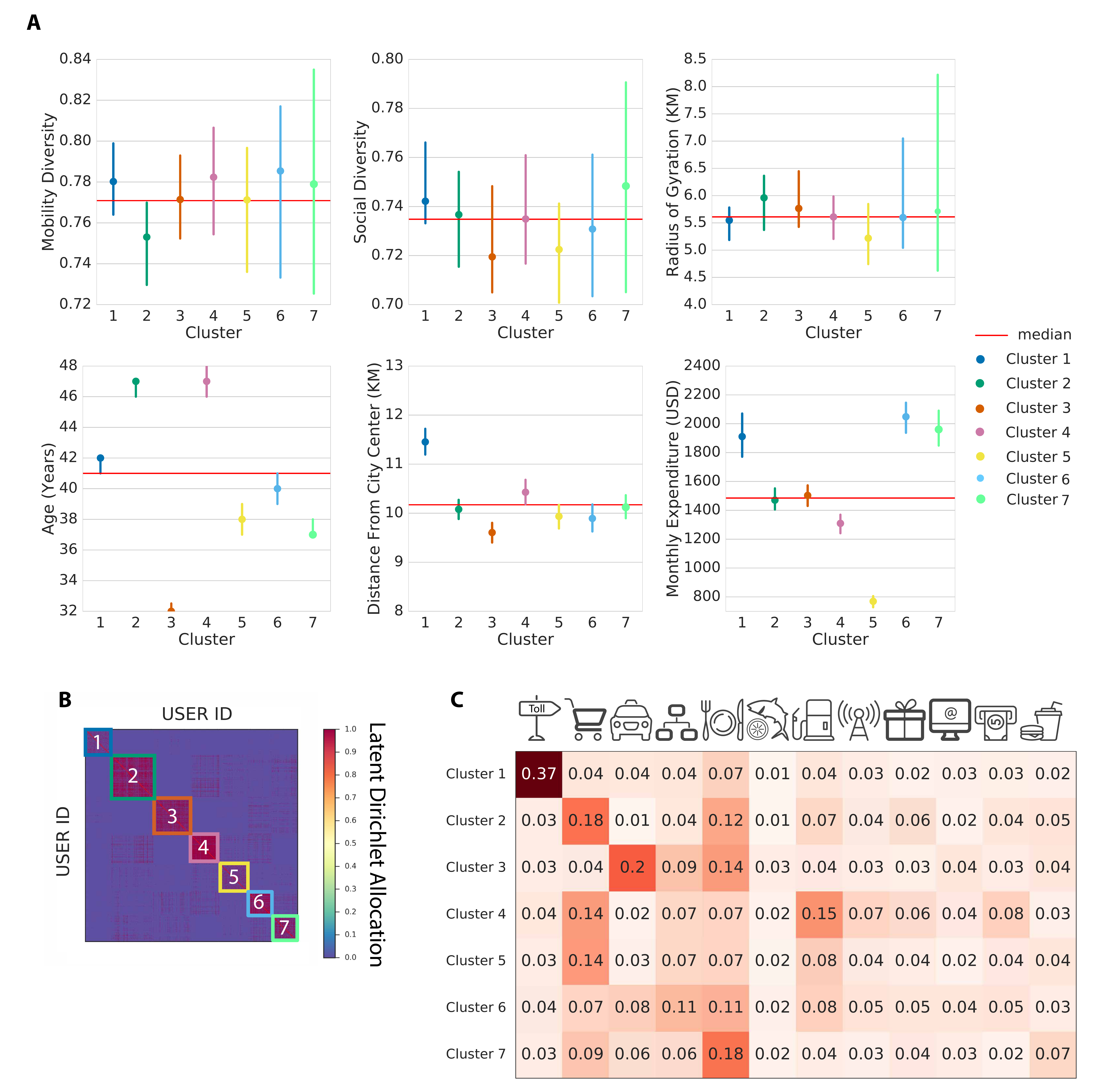}
\caption{Cluster analysis of the selected user using Latent Dirichlet Allocation (LDA) \cite{blei2003latent} to model user transactions. \textbf{(A)} Confidence interval of $95\%$ of socio-demographic characteristics of the seven clusters detected the solid in red representing the median values of the all clustered users. \textbf{(B)} Cluster analysis of the 13.0K selected users. Using LDA we model each user as a mixture of five spending behaviors, where each behavior is a mixture of transaction codes. We compute the Jensen-Shannon divergence \cite{lin1991divergence} for the user similarity matrix, then perform the Louvain algorithm with a threshold of 0.1. We compare the clusters detected with the LDA and the Sequitur methods using the Normalized Mutual Information \cite{danon2005comparing,ana2003robust} NMI = 0.2 and the Rand=0.7 \cite{rand1971objective}. This two tests show a degree of similarity among the clusters. \textbf{(C)} Frequency of the transaction codes for the 10 weeks considered for each of the seven clusters detected. The clusters extracted with the LDA manifest similar characteristics with the ones extracted with our method (Supplementary Figure 13B). The (1,2,3,7) clusters detected by LDA share distribution in spending codes with the (1,2,3,6) cluster of the sequitur. Moreover, the clusters (4,5) of LDA are very similar at the sequitur cluster 5. The icons used in this figure are work of Azaze11o/Shutterstock.com.}
\label{fig:Sfig18}
\end{figure}

\begin{figure}[!htb]
\centering
\includegraphics[width=\linewidth]{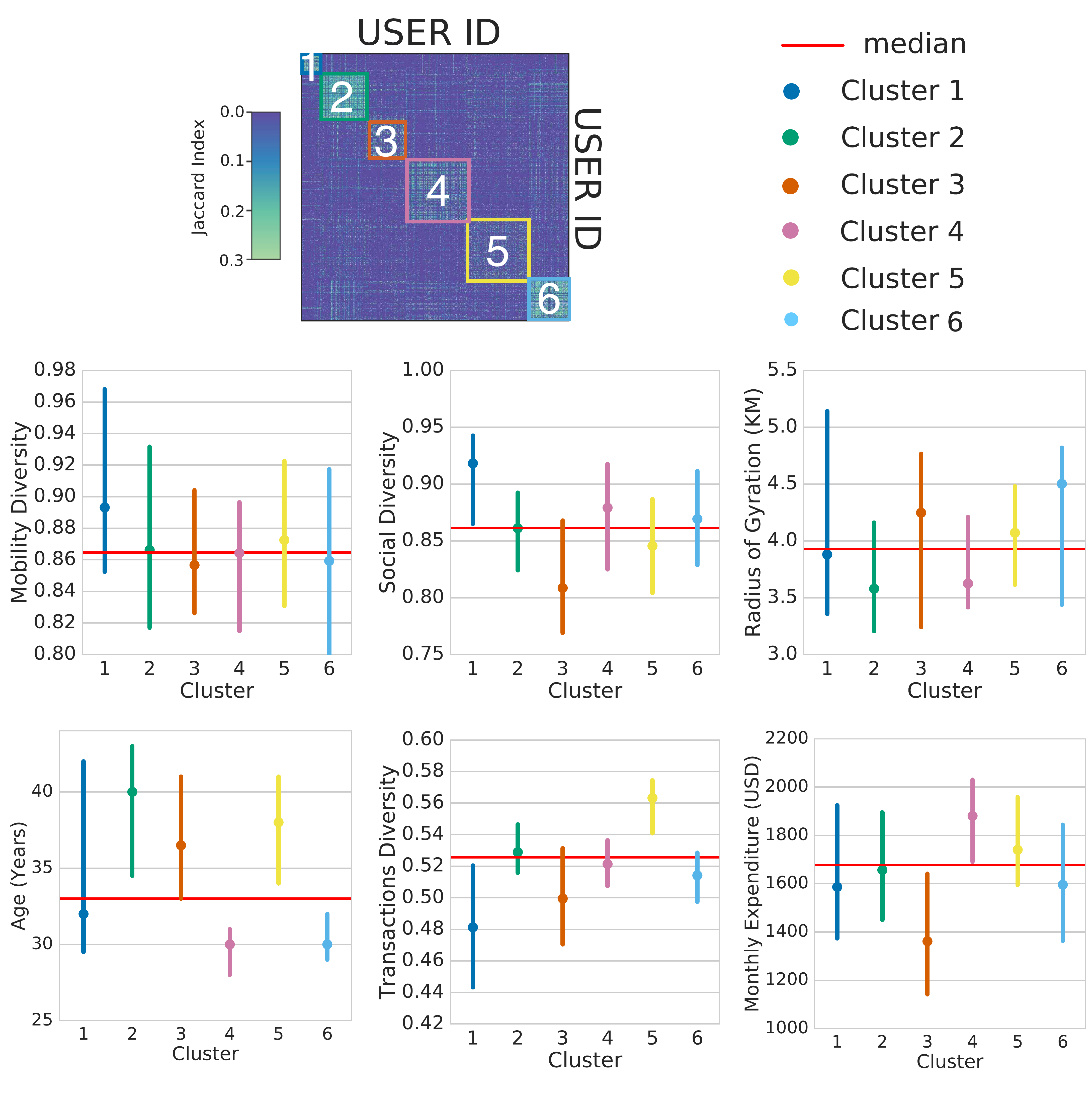}
\caption{ \textbf{Cluster analysis for the second city analyzed in Mexico: Puebla.} Confidence interval of $95\%$ of socio-demographic characteristics. In this city the users have an higher mobility and social diversity with low radius of gyration. (For further considerations see Supplementary Figure 20-21)}
\label{fig:Sfig19}
\end{figure}

\begin{figure}[!htb]
\centering
\includegraphics[width=0.83\linewidth]{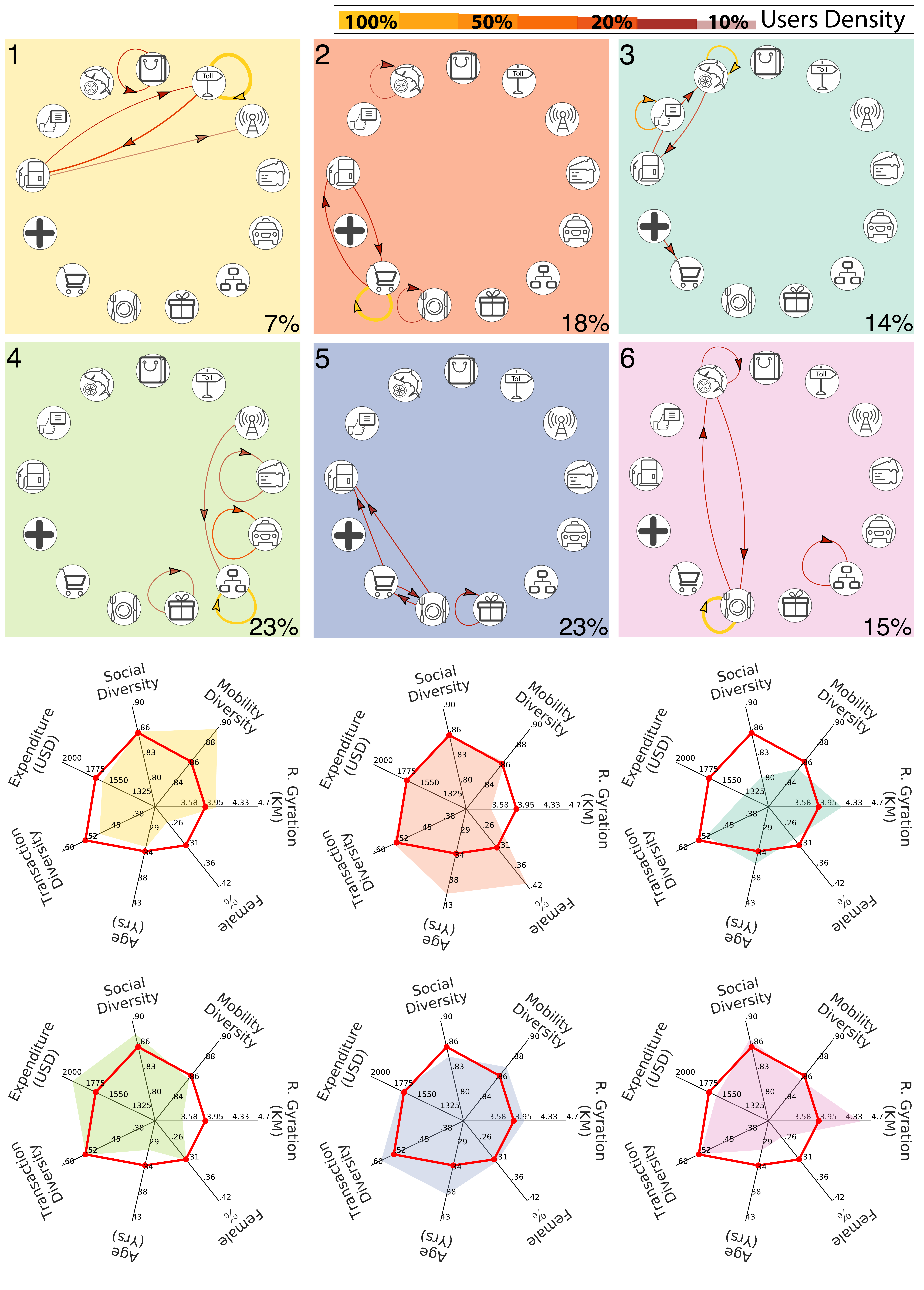}
\caption{ \textbf{Groups based on their spending habits for the second city analyzed in Mexico: Puebla.} We show the top 5 most frequent spending sequences of the users in each group, representing more than $30\%$ of users' shopping routines. The percentage of the total users in each group is shown in the bottom-right corner. Distribution of individual characteristics among users: gender radius of gyration, mobility diversity, social diversity, median expenditure by month, transaction diversity and age. While the clusters (1,2,4,5,6), manifest similarity among the two cities. The cluster 3 in the City B has different routines with the core transactions in Miscellaneous Food store and insurance instead of taxi and restaurants (see Supplementary Figure 21 for further comparison on the socio-demographic-mobility indicators between the two cities). The icons used in this figure are work of Azaze11o/Shutterstock.com.}
\label{fig:Sfig20}
\end{figure}

\begin{figure}[!htb]
\centering
\includegraphics[width=0.8\linewidth]{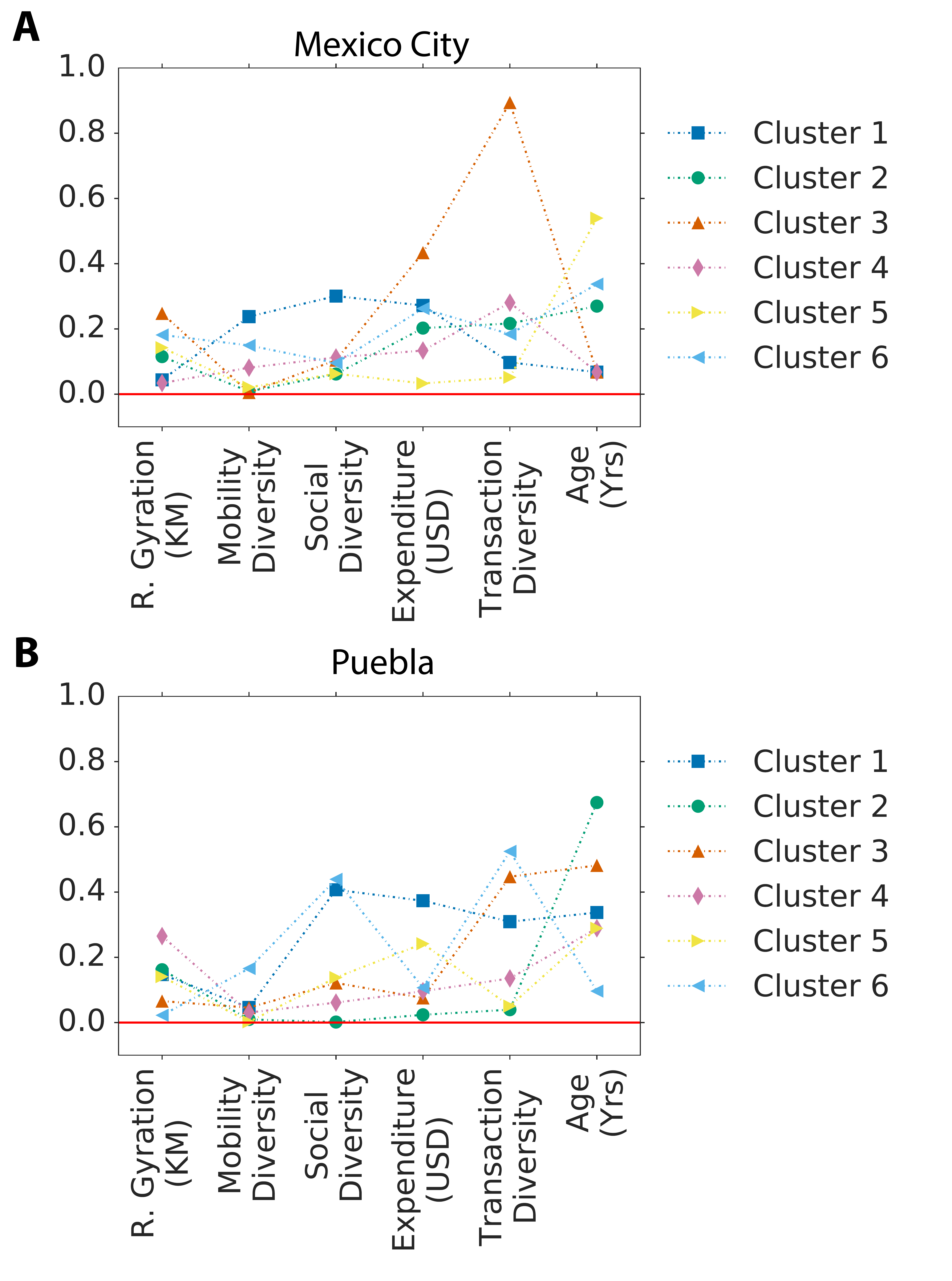}
\caption{ \textbf{Analysis of median socio-demographic-mobility index variation per cluster}. The y axis represents $(\tilde{x_{i}}-\tilde{X})/\text{MAD}(X)$; with $\tilde{X}$ the median of the whole dataset for the socio-demographic-mobility attribute $X$, $\tilde{x_{i}}$ the median of the same socio-demographic-mobility attribute of the $i$-cluster users and MAD the Median Absolute Deviation. Remarkably the behaviors of the clusters (2,4,5,6) are very similar between the two cities considered. The two Clusters 3 as already stress represent two different segments of the population. Meanwhile the clusters 1 of the commuters have different behaviors maintaining the lower transaction diversity this could be due to the different topology of the cities.}
\label{fig:Sfig21}
\end{figure}

\end{document}